\documentclass[12pt]{article}
\pdfoutput=1
\usepackage{cite}
\usepackage{graphicx}
\usepackage{multicol}
\usepackage{amsfonts}
\usepackage{amssymb}
\usepackage{amsmath}
\usepackage{heck}
\usepackage{setspace}
\usepackage[all]{xy}
\usepackage{verbatim}
\usepackage{color}
\usepackage{float}
\usepackage{epsfig}
\usepackage{multirow}
\usepackage{cancel}
\usepackage{mcite}
\usepackage{youngtab}
\usepackage{epstopdf}
\usepackage{wrapfig}

\newcommand{\ba}{\begin{eqnarray}}
\newcommand{\ea}{\end{eqnarray}}

\begin{document}

\def\simgt{\mathrel{\lower2.5pt\vbox{\lineskip=0pt\baselineskip=0pt
           \hbox{$>$}\hbox{$\sim$}}}}
\def\simlt{\mathrel{\lower2.5pt\vbox{\lineskip=0pt\baselineskip=0pt
           \hbox{$<$}\hbox{$\sim$}}}}

\begin{titlepage}

\begin{flushright}
\end{flushright}

\begin{center}

{\Large \bf ``$L=R$" -- $U(1)_R$ as the Origin of Leptonic `RPV'}

\vskip 1cm

{\large Claudia Frugiuele$^{\dag,c}$, Thomas Gr\'egoire$^{\dag}$, Piyush Kumar$^{*,\S}$, Eduardo Pont\'on$^*$}

\vskip 0.4cm

$^{\dag}${\it Ottawa-Carleton Institute for Physics \\ Department of Physics, Carleton University \\
1125 Colonel By Drive, Ottawa, K1S 5B6 Canada}

\vskip 0.4cm

$^{c}${\it Theoretical Physics Department\\ Fermilab \\
P.O. Box 500, Batavia, IL 60510 USA}

\vskip 0.4cm

$^*${\it Department of Physics $\&$ ISCAP   \\ 538 W 120th street \\
Columbia University, New York, NY 10027 USA}

\vskip 0.4cm

$^{\S}${\it Department of Physics \\
Yale University, New Haven, CT 06520 USA}

\abstract{A classification of phenomenologically interesting
supersymmetric extensions of the Standard-Model with a $U(1)_R$
symmetry is presented.  Some of these are consistent with subsets of
leptonic or baryonic ``R-parity violating" (RPV) operators, thereby
providing a natural motivation for them.  We then focus on a
particular class of models in which the $U(1)_R$ symmetry coincides
with lepton number when restricted to the SM sector.  In this case,
the extension of lepton number to the superpartners is
``non-standard", implying, in particular, the existence of the
leptonic RPV operators $LLE^c$ and $LQD^c$,
and a vacuum structure where one of the left-handed sneutrinos
acquires a significant vacuum-expectation-value, while not being
constrained by neutrino mass bounds.  The model can be naturally
consistent with bounds from electroweak precision measurements and
flavor-changing processes.  It can also easily accommodate the
recently measured Higgs mass due to the existence of a scalar triplet
that couples to the Higgs with an order one coupling, with only
moderate fine-tuning.  The phenomenology is rather rich and
distinctive, with features such as heavy-but-natural Dirac gauginos,
relaxed bounds on squarks, resonant slepton/sneutrino production,
lepto-quark signals, as well as an interesting connection to neutrino
physics arising from $R$-breaking.  The broad qualitative features are
discussed in this paper, with a more detailed phenomenological study
carried out in a companion paper \cite{Frugiuele:2012xx}.}

\end{center}
\maketitle

\end{titlepage}
\tableofcontents


\section{Introduction}
\label{intro}

We are extremely fortunate to live in a data-rich era of particle
physics.  The discovery of a Higgs-like particle
\cite{:2012gu,:2012gk} is indeed a monumental
achievement of the LHC. Measuring the properties of this particle in
detail is now one of the most important experimental tasks.  On the
other hand, the mass of this particle ($\sim 125$ GeV), as well as
null results for beyond the Standard-Model (SM) physics so far, have
started challenging our simple expectations for physics beyond the
SM.\footnote{For concreteness, we interpret the $\sim 125$ GeV
resonance as arising from a CP-even Higgs-like particle.} In
particular, the Minimal Supersymmetric Standard Model (MSSM), which is
the leading candidate for beyond-the-Standard Model (BSM) physics, is
being significantly constrained.  For example, the bounds
on colored superpartners are well over a TeV in the ``bulk" of
parameter space, where pair production of squarks and/or gluinos is
followed by decay into quarks or gluons and the lightest
supersymmetric particle (LSP), giving rise to a jets plus
$\slash{\!\!\!\!E}_T$ signature.

How can one interpret the Higgs-like discovery at $\sim$125 GeV, and
the null results for new physics so far?  Our discussion will be
within the supersymmetric paradigm for concreteness.  The current data
seems to suggest two rather different approaches to beyond-the-SM
(BSM) physics.  One viewpoint is that a Higgs near 125 GeV and the
absence of superpartners so far, can be explained in the MSSM with
heavy ($\sim 10-100$ TeV) scalars.  These models are therefore
\emph{electroweak-tuned}.  Models of this type can arise in some
versions of \cite{Wells:2004di, ArkaniHamed:2004fb}.  In fact, there
exist top-down frameworks which predict heavy scalars in the 10-100
TeV range \cite{Acharya:2007rc, Acharya:2008zi} and, therefore, a
Higgs mass near the observed value \cite{Kane:2011kj}; the \emph{big}
hierarchy problem of dynamically generating the ($\sim$10) TeV scale
from the Planck scale is solved in such a framework, but a
\emph{little} hierarchy remains.

However, the other viewpoint is that it is premature to completely
abandon \emph{electroweak naturalness}\footnote{It is hard to give a
precise unambiguous criterion of electroweak-naturalness, but the
general notion is rather clear.} at this stage, for it is still
possible to imagine models in which the bounds on superpartners are
evaded in a natural manner, allowing {\it electroweak-natural} (at
least to a large degree) models.  At the same time, various mechanisms
exist which can give rise to a Higgs-like particle at around 125 GeV
without introducing excessive tuning.  With this philosophy in mind,
some approaches to (SUSY) BSM physics have generated renewed interest,
such as: i) Models in which the first and second generation squarks
are rather heavy, but those of the third generation are light for some
reason, as
in~\cite{Cohen:1996vb,Sundrum:2009gv,Gherghetta:2011wc,Csaki:2012fh,
Larsen:2012rq,Cohen:2012rm,Craig:2012di}, ii) Models which have a
rather special spectrum, such as a compressed one
\cite{LeCompte:2011fh, LeCompte:2011cn,Murayama:2012jh}, or a
stealth one \cite{Fan:2011yu, Fan:2012jf}, iii) Models which do not
give rise to signatures with appreciable missing energy, the
prototypical example being that of baryonic $R$-parity violating (RPV)
models (see \cite{Barbier:2004ez} for a review), and iv) Models in
which the production cross-section is small even for light first and
second generation squarks, as in examples with a relatively heavy
Dirac gluino \cite{Kribs:2012gx,Heikinheimo:2011fk}.

In \cite{Frugiuele:2011mh}, a model was proposed that shares some
features of the RPV models and models with Dirac gluinos, mentioned
above.  The defining feature of this model was the existence of a
$U(1)_R$ symmetry that was identified with one of the lepton numbers
and the role of the sneutrino as the down-type Higgs.  In
\cite{Bertuzzo:2012su} the R-symmetry was generalized to a global
lepton number to allow for viable neutrino masses and mixings.
Ref.~\cite{Brust:2011tb} studied the case where the $U(1)_R$ symmetry
is identified with the baryon number.  In this paper we perform a
classification of phenomenologically interesting $R$-symmetric models,
and show how some of these $R$-symmetries are consistent with leptonic
or baryonic RPV operators, and therefore provide an elegant motivation
for the existence of these operators (see \cite{Csaki:2011ge} for an alternative approach motivating RPV operators).  
We then focus on the case where the lepton number is tied to $U(1)_R$ symmetry, and study it in detail
with the goal of laying out the LHC phenomenology.

Indeed, having an $R$-symmetry gives rise to many interesting
phenomenological features.  For example, Majorana gaugino masses,
certain scalar trilinear ``$A$-terms", and the ``$\mu$-term" are
forbidden.  However, gaugino masses of the Dirac type are
allowed.\footnote{Dirac gaugino masses can also be motivated from
``supersoft" supersymmetry breaking in which the gauge sector has
${\cal N}=2$ supersymmetry \cite{Fox:2002bu}.} This leads to a
significant suppression of flavor and CP-violating effects relative to
the MSSM for ${\cal O}(1)$ flavor-violating soft scalar masses and
phases \cite{Kribs:2007ac}.  Various other aspects of $R$-symmetric
models have been studied in the
literature~\cite{Hall:1990dga,Hall:1990hq,Nelson:2002ca,Chacko:2004mi,Benakli:2008pg,Choi:2008pi,Kumar:2009sf,
Dobrescu:2010mk,Benakli:2011vb,Davies:2011js}.  Although the minimal
$R$-symmetric spectrum does not give rise to gauge coupling
unification, many scenarios for adding additional matter have been
proposed which could help unify the couplings \cite{Fox:2002bu,
Benakli:2010gi, Abel:2011dc}.  Particular variants of $R$-symmetric
models can also give rise to a strong electroweak phase transition
generating the observed baryon asymmetry, as well as an LSP DM
candidate \cite{Kumar:2011np, Fok:2012fb}.  Finally, a remarkable
feature, which we will exploit in this work, is that when the
$R$-symmetry is identified with a lepton or baryon number
respectively,\footnote{The lepton and baryon numbers of the SM
fermions are standard, but the extension of these to BSM particles is
``non-standard".} this allows leptonic or baryonic RPV operators in
the Lagrangian consistent with these symmetries.  Therefore, these are
not subject to stringent constraints from lepton or baryon number
violating observables, such as upper bounds on neutrino masses and
nucleon-antinucleon oscillations, respectively.

$R$-symmetries are also well-motivated from a more theoretical point
of view.  In the global supersymmetric limit, it has been known for
quite some time that $R$-symmetry plays an important role in
supersymmetry breaking as it is directly related to the existence of
supersymmetry breaking minima due to the Nelson-Seiberg theorem
\cite{Nelson:1993nf}.  Since Majorana gaugino masses necessarily break
the $R$-symmetry, considerable effort has been devoted to generating
large-enough Majorana gaugino masses while still preserving enough
$R$-symmetry to keep supersymmetry breaking intact.  However, an
alternative is to consider generic supersymmetry breaking scenarios
that give rise to \emph{Dirac} gaugino masses fully consistent with
the $R$-symmetry.

The scope of this work is the following.  We present a classification
of phenomenologically viable $R$-symmetric models by providing a
rather general description of such models in Section~\ref{models}.
This is done by making manifest the relevant $U(1)$ symmetries
present.  Then, we show that preserving different combinations of
$U(1)$ symmetries gives rise to the different variants of
$R$-symmetric models considered in the literature, including some
which have been relatively poorly explored.  In particular, we will
see that in addition to the relatively well-studied $R$-symmetric
models with the usual ``$R$-parity conserving" operators, there are
models which include subsets of the so-called ``$R$-parity violating
(RPV)" operators.  However, it is important to note that while the
standard $R$-parity is violated in these models, there is still a
continuous $R$-symmetry and these operators are perfectly compatible
with it.  These kinds of $R$-symmetric models arise when one
identifies the lepton number ($L$) or the baryon number ($B$) of the
SM fermions with their $R$-charges, as will be clear soon.  When the
lepton number behaves as an $R$-symmetry, depending upon the region of
parameter space, it is possible to have either the usual down-type
Higgs ($H_d$) providing masses to the down-type fermions, or have one
of the sneutrinos ($\tilde{\nu}_1$) providing such masses since it
also gets a vacuum-expectation value ({\it vev}).  Furthermore, in the
limiting case when $H_d$ is heavy and is not part of the low-energy
spectrum, then the dominant contribution to down-type masses arises
from the sneutrino {\it vev}, which is \emph{not} constrained by
neutrino masses unlike that in standard RPV models.  All these points
will be explained in detail in the following sections.  After
describing the general classification of models in
Section~\ref{models}, in the rest of the paper we specialize to the
case where the lepton number is related to the $R$-symmetry, as
suggested by the title of the paper.  We present some characteristic
features of the model in Section~\ref{charac}, highlighting the
differences from standard RPV models.  The fermionic electroweak
sector of the model is studied in Section~\ref{fermionic-ew}, followed
by a discussion of the existing constraints on the model from indirect
effects, such as electroweak precision observables, flavor physics,
etc.  in Section~\ref{sec:ewp}.  The basic aspects of the Higgs sector
of these models are laid out in Section~\ref{higgs}, in particular the
region of parameter space which gives rise to a $\sim$125 GeV CP-even
eigenstate is explained.  Finally, we describe the broad
phenomenological features of the model in Section~\ref{pheno}.  Since
the collider signals of these scenarios are quite interesting and
novel, in this paper we discuss only the qualitative features which
set them apart from other models.  A more detailed treatment of
collider constraints and signals is done in a companion paper
\cite{Frugiuele:2012xx}.  Finally, we briefly discuss some aspects of
Dark Matter in Section~\ref{dm}, followed by conclusions and future
directions in Section~\ref{conclude}.  The appendices deal with some
details of $R$-symmetry breaking operators in
Appendix~\ref{r-breakops}, a description of a flavor ansatz for the
$\lambda$ and $\lambda'$ couplings (the standard notation for the
coefficients of the $LLE^c$ and $LQD^c$ operators, respectively) in
Appendix~\ref{ansatz}, and an estimate of the lower bound on
$\lambda'_{i33}$ couplings, given the observation of a lepto-quark
(LQ) signal, in Appendix~\ref{lambdap-bound}.

\section{Classification of $R$-symmetric Models}
\label{models}

We start with the prototypical $R$-symmetric Standard Model studied in
\cite{Kribs:2007ac}.  In addition to the superfield content of the
MSSM -- $H_{u}$, $H_{d}$, $Q_{i}$, $U^{c}_{i}$, $D^{c}_{i}$, $L_{i}$,
$E^{c}_{i}$ ($i=1,2,3$), this includes a pair of vector-like $SU(2)_L$
doublet superfields $R_u$ and $R_d$ (with hypercharge $\frac{1}{2}$
and -$\frac{1}{2}$ respectively), as well as superfields in the
adjoint representation of the SM gauge group: a ``hypercharge adjoint" or
singlet, $S$, an $SU(2)_{L}$ triplet, $T$ (with zero hypercharge), and
an $SU(3)_{C}$ octet, $O$.
 
The relevant global $U(1)$ symmetries of the model in
\cite{Kribs:2007ac} include an $R$-symmetry denoted by $U(1)_{R_0}$
along with the well known lepton number $U(1)_L$ and baryon number
$U(1)_B$, as shown below in Table~\ref{charges-old}:
\vspace{-4mm}
%
\begin{table}[h!]
\begin{center}
\begin{tabular}{|c|cccccccccccc|}
\hline
\rule{0mm}{5mm}
& $Q_{i}$ & $U_{i}^{c}$ & $D_{i}^{c}$ & $L_{i}$ & $E_{i}^{c}$  & $H_{u}$ & $H_{d}$ & $R_{u}$ & $R_{d}$ & $S$ & $T$ & $O$ 
\\[0.3em]
\hline
\rule{0mm}{5mm}
$U(1)_{R_0}$ & 1 & 1 & 1 & 1 &1& 0 & 0 & 2 & 2 & 0 & 0& 0\\
[0.3em]
$U(1)_{L}$ & 0 & 0 & 0 & 1 & -1 & 0 & 0 & 0 & 0 & 0 & 0 & 0\\
[0.3em]
$U(1)_{B}$ & 1/3 & -1/3 & -1/3 & 0 & 0 & 0 & 0 & 0 & 0 & 0 & 0 & 0\\
[0.3em]
\hline
\end{tabular}
\end{center}
\vspace{-10pt}
\caption{(Super)Field content and $U(1)$ charge assignments.}
\label{charges-old}
\vspace{-15pt}
\end{table}

The following superpotential consistent with the above symmetries was
considered in~\cite{Kribs:2007ac}:
\ba W_0 = y^{u}_{ij} H_{u} Q_{i} U^{c}_{j}  
+ \tilde{y}^{d}_{ij} H_d Q_{i} D^{c}_{j} + \tilde{y}^{e}_{ij} H_{d} L_{i} E^{c}_{j}~ 
+ \mu_u H_{u} R_{d} +   \mu_d R_{u} H_{d}~.\label{W-KPW} 
\ea 
It is possible to also write down the following terms with adjoint
superfields, consistent with all symmetries:
\ba W_{adj} = S (\lambda^{S}_u H_{u} R_{d}  + \lambda^{S}_d R_{u} H_{d}) + 
(\lambda^{T}_u H_{u} T  R_{d} + \lambda^{T}_d R_{u}  T H_{d})~. \label{W-adj}
\ea 
It is easy to see that with the above $R$-charge assignments the usual
``RPV" operators, schematically denoted by $L H_u$, $L L E^c$, $L Q
D^c$, and $U^c D^c D^c$, are forbidden.  However, $R$-symmetries are
\emph{not} inconsistent with subsets of RPV operators in general,
since it is possible to construct $R$-symmetries $R_i$ ($i=1,2,3$), which
are linear combinations of $U(1)_{R_0}$, $U(1)_L$ and $U(1)_B$, as
seen in Table \ref{charges-new} below:
%
\begin{table}[h!]
\begin{center}
\begin{tabular}{|c|cccccccccccc|}
\hline
\rule{0mm}{5mm}
& $Q_{i}$ & $U_{i}^{c}$ & $D_{i}^{c}$ & $L_{i}$ & $E_{i}^{c}$  & $H_{u}$ & $H_{d}$ & $R_{u}$ & $R_{d}$ & $S$ & $T$ & $O$ 
\\[0.3em]
\hline
\rule{0mm}{5mm}
$U(1)_{R_1=R_0-L}$ & 1 & 1 & 1 & 0 & 2& 0 & 0 & 2 & 2 & 0 & 0& 0\\
[0.3em]
$U(1)_{R_2=R_0+B}$ & 4/3 & 2/3 & 2/3 & 1 & 1 & 0 & 0 & 2 & 2 & 0 & 0 & 0\\
[0.3em]
$U(1)_{R_3=R_0+L}$ & 1& 1 & 1 & 2 & 0 & 0 & 0 & 2 & 2 & 0 & 0 & 0\\
[0.3em]
\hline
\end{tabular}
\end{center}
\vspace{-15pt}
\caption{(Super)Field content and three different combinations of
$U(1)_R$ charge assignments.}
\label{charges-new}
\vspace{-10pt}
\end{table}

We see that depending on the choice of the $R$-symmetry, the following
RPV operators are allowed in the superpotential, {\it in addition} to
those in (\ref{W-KPW}) and (\ref{W-adj}):
\ba\label{WLR}
W_1 &=& W_0 + W_{adj} +  \lambda_{ijk} L_{i} L_{j} E^{c}_{k} +\lambda'_{ijk} L_{i} Q_{j} D^{c}_{k}~,
\hspace{1.5cm} (R_1=R_0-L)
\nonumber \\ [0.3em]
W_2 &=& W_0 + W_{adj} +  \lambda''_{ijk} U^c_{i} D^c_{j} D^{c}_{k}~,
\hspace{3.9cm} (R_2=R_0+B) 
\\ [0.3em]
W_3 &=&  W_0 + W_{adj} + \mu^{(i)}_L H_u L_i~.
\hspace{4.6cm}(R_3=R_0 + L)
\nonumber
\ea
In principle, the term $\mu_L^{(i)} R_u L_i$ is also allowed for
$R=R_1$, but it is possible to do an $SU(4)$ field redefinition of the
$L_i$ and $H_d$, and define $``H_d"$ as the field which couples to
$R_u$, leading to (\ref{WLR}).

Thus, this provides a rather general classification of
phenomenologically viable $R$-symmetric Standard Models.  The choice
$R=R_1$ corresponds to identifying the lepton numbers of the SM
fermions with (the negative of) their $R$-charges, while the choice
$R=R_3$ corresponds to identifying them with their $R$-charges.  The
choice $R=R_2$ on the other hand identifies the baryon numbers of the
SM fermions with their $R$-charges, and has been considered in
\cite{Brust:2011tb}.  Note that since the $R$-symmetries $R_{1,2,3}$
are identified with lepton or baryon number, they are anomalous and
should therefore be thought of as accidental low-energy symmetries
just like the latter (note that $R_0$ is non-anomalous).  The above
are special limits of a generic $U(1)_{\hat{R}}$ symmetry with
$\hat{R} = R + a L + b B$ for real $a$ and $b$.  However, all other
cases are severely constrained by proton decay bounds, and we do not
consider them any further.

As we will see later, even though the $R$-symmetries $ R_1, R_2$ and
$R_3 $ allow RPV operators, they are less constrained than standard
RPV models with the same operators.  The basic reason for this is that
in these models these RPV operators are consistent with lepton or
baryon numbers (which are identified with the $R$-symmetries above),
hence there are no constraints on the couplings of these operators
from processes which violate lepton or baryon number.

\subsection{Supersymmetry Breaking}
\label{susy-break}

In order to fully specify the lagrangian, supersymmetry breaking terms
must be included.  Since we are interested in $R$-symmetric models, we
imagine a situation in which supersymmetry breaking (at least in the
global limit) is \emph{not} accompanied by $R$-breaking.  This can
happen if supersymmetry breaking is of the \emph{$D$-type}, as 
described in \cite{ArkaniHamed:2004yi}.  This includes both $D$-term
SUSY breaking parametrized by a spurion superfield ${\cal W}'_{\alpha}
= \lambda'_{\alpha}+\theta_{\alpha} D'$ with $R[{\cal W}'_{\alpha}]=1$
($R[\lambda'_{\alpha}]=1, R[D'] = 0$) and $\langle D' \rangle \neq 0$,
as well as $F$-term supersymmetry breaking parametrized by a spurion
superfield $X =x+\theta^2\,F_X$ with $R[X]=2$ and $\langle x \rangle
=0,\,\langle F_X \rangle \neq 0$.

We follow the same procedure as before, by first considering soft
terms consistent with the original $R$-charge assignments ($R=R_0$) as
in Table \ref{charges-old}, and then studying additional terms allowed
by the other choices - $\{R_1, R_2, R_3\}$ in Table \ref{charges-new}.

The spurion ${\cal W}'_{\alpha}$ generates ``supersoft''
terms via\footnote{The ${\cal W}^{(i)}_{\alpha}$ are the $SU(3)_{C} \times
SU(2)_{L} \times U(1)_{Y}$ chiral superfield strengths.} 
\ba
\sqrt{2} \int \! d^{2}\theta \, \frac{{\cal W}^{\prime \alpha}}{M_{\star}} 
\left[ c_{1} {\cal W}^{(1)}_{\alpha} S + c_{2} {\cal W}^{(2)i}_{\alpha} T^{i} + 
c_{3} {\cal W}^{(3) a}_{\alpha} O^{a}  \right] + {\rm h.c.}~,
\label{supersoft}
\vspace{-10pt}
\ea
\noindent which contain Dirac gaugino masses $m_{D_{i}} = c_{i}
D'/M_{\star}$.  Here $M_{\star}$ denotes the scale of SUSY breaking
mediation (e.g.~the messenger scale in Dirac gauge-mediation
scenarios).  The above terms preserve a $U(1)_{R}$ symmetry under
which the ${\cal W}^{(i)}_{\alpha}$ and ${\cal W}'_{\alpha}$ have
$R$-charge\footnote{The choice $R[{\cal W}_{\alpha}]=R[{\cal
W}'_{\alpha}]=1$ is dictated by the SUSY gauge kinetic terms; in
particular this always implies
$R[\lambda_{\alpha}]=R[\lambda'_{\alpha}]=1$.} 1, while $R[S] =
R[T^{i}] = R[O^{a}] = 0$.

The spurion $X$ generates the following $U(1)_{R}$-preserving
renormalizable soft terms:
\ba
L^{\rm soft}_0 &=& \sum_{i} m^{2}_{i} \Phi_{i}^{\dagger} \Phi_{i} 
+ \left[t_{S} S + \frac{1}{2} b_{S} S^{2} + \frac{1}{3} A_{S} S^{3} + 
\frac{1}{2} b_{T} T^{2} + \frac{1}{2} b_{O} O^{2} + B\mu\, H_u H_d
\right.
\nonumber\\
&& \hspace{2.6cm}
\mbox{} + \left. \rule{0mm}{6mm}
A_S S H_u H_d +A_T H_u T H_d +A^{\lambda}_T S T^2 + A^{\lambda}_O S O^2 + {\rm h.c.} \right] ,
\label{soft-R}
\ea
where the sum runs over all the scalars, and we denote the scalar
components by the same notation used for the superfields.  These are
generated via the following operators:
\ba
\label{softops}
&& \hspace*{-1cm}
\int \! d^{4} \theta \, \frac{X^{\dagger} X}{M_{\star}^{2}} \left\{ 
\sum_{i} \Phi^{\dagger}_{i} \Phi_{i} +  \left[ H_u H_d + \epsilon\,M_{\star} S + S^{2} + T^{2} + O^{2} + \frac{1}{M_{\star}} \times \textrm{cubic} + {\rm h.c.} \right]
\right\}~,
\\[0.5em]
&& \hspace*{-1cm}
\int \! d^{2} \theta \, \frac{X}{M_{\star}}  (S T^2 + S O^2 + S^{3}) + {\rm h.c.}~.
\label{Wtrilinear}
\ea
We can see that operators quadratic in
the visible superfields in the first line in (\ref{softops}) are of
order $\frac{|F_X|^2}{M_{\star}^2}$.  We take $F_{X} \sim D'$ and
\ba
\frac{F_{X}}{M_{\star}} &\equiv& M_{\rm SUSY} ~\sim~ 100~{\rm GeV}-1~{\rm TeV}~. 
\ea 
So Dirac gaugino masses from (\ref{supersoft}), and the
non-holomorphic soft mass-squareds, and $B$-terms from (\ref{softops})
are parametrically of the same order (although there may be modest
numerical hierarchies).  The operators which are cubic and higher
order in the visible superfields in Eq.~(\ref{softops}) will be
suppressed by powers of $M_{\rm SUSY} / M_{\star}$, and are therefore
very suppressed.  For the linear term in $S$ in (\ref{soft-R}),
dimensional analysis generically gives a coefficient $t_S$ of order
$M_{\star}\,M_{{\rm SUSY}}^2$.  Phenomenologically however, $t_{S}$
should not be larger than $M_{\rm SUSY}^{3}$, since otherwise the
scalar singlet tadpole, $t_{S}\,S$, will destabilize the hierarchy.
Ref.~\cite{Goodsell:2012fm} has recently argued that this is indeed
the case in these scenarios, so that one has $\epsilon \ll 1$.

The operators in the second line in (\ref{Wtrilinear}) can give
trilinear $A$-terms involving the adjoint fields of order
$\frac{|F_X|}{M_{\star}} \sim M_{{\rm SUSY}}$ if allowed.  However,
these are forbidden if $X$ is not a gauge singlet, implying that the
scale of these operators can be easily suppressed relative to those in
the first line, whose scale is naturally set by $M_{\rm SUSY}$.
Finally, note that since we imagine that $X$ belongs to a hidden
sector which has no direct couplings to the observable sector
superfields above, there are no terms like $\int \!  d^{2} \theta \, X
[ M_{\star} S + H_u\,H_d+S^{2} + T^{2} + O^{2}]$ (due to the
non-renormalization theorem, it is technically natural to omit these
superpotential couplings).

With other choices for the $R$-symmetry - $R_1$, $R_2$ and $R_3$, as
described in Table \ref{charges-new}, one can write down additional
soft supersymmetry breaking operators as follows:
\ba 
L^{\rm soft}_1 &=& L^{\rm soft}_0 + B\mu_L^{(i)} H_u L_i +  
A_{S}^{(i)} S H_{u} L_{i} + A_{T}^{(i)} H_{u} T L_{i}~,
\hspace{0.75cm} R_1 = R_0-L
\nonumber \\ [0.3em]
L^{\rm soft}_2 &=& L^{\rm soft}_0,\hspace{8.05cm} R_2 = R_0+B 
\label{soft-new}
\\ [0.3em]
L^{\rm soft}_3 &=& L^{\rm soft}_0~.
\hspace{8cm} R_3 = R_0+L
\nonumber
\ea 
Only the case $R_1= R_0-L$ allows additional gauge-invariant operators
consistent with the $R$-symmetry (since $L$ has zero $R$-charge).
These additional soft terms for the case $R=R_1$ in (\ref{soft-new})
give rise to a very interesting possibility for the vacuum structure
of the theory.

The presence of the $B\mu_L^{(i)}$ term in (\ref{soft-new}) for
$R=R_1$ implies that one of the left-handed sneutrinos gets a {\it
vev}~\footnote{Recall that we have defined ``$H_d$" as the linear
combination of $SU(2)_L$ doublets with $Y = -1/2$ and $R = 0$, that
couples to $R_u$ in the superpotential.  Within this class of bases,
it is further possible, by $SU(3)$ rotations, to go to the
``single-vev-basis" where only one of the three $L_i$ acquires a
\textit{vev} (see Section~\ref{charac}).  We will see later that the
most natural choice is to identify the direction of this {\it vev}
with the ``electron" direction, i.e.~$i=1$.  The $H_d$ \textit{vev} in
this basis may be non-vanishing, but we are interested in a region of
parameter space where it is small compared to the EW scale.  } due to
a tadpole for the sneutrino when $H_u$ gets a {\it vev}.  It then
becomes possible to distinguish two extreme cases: i) $\langle
\tilde{\nu}_1\rangle \ll \langle H_d^0 \rangle$, and ii) $\langle
\tilde{\nu}_1\rangle \gg \langle H_d^0 \rangle$.  In fact the size of
the $\mu_d\,R_u H_d$ term in superpotential $W_0$ controls which one
is relevant.  This is because, schematically,
\ba 
\langle \tilde{\nu}_1\rangle &\sim& \frac{B\mu_L^{(1)}}{m_{\tilde{L}}^2}\,v_u~,
\hspace{2mm}
\langle H_d^0 \rangle \sim \frac{B\mu}{\mu_d^2}\,v_u
\hspace{1cm}
\implies 
\hspace{1cm}
\frac{ \langle H_d^0 \rangle}{ \langle \tilde{\nu}_1 \rangle} ~\sim~ \frac{B\mu}{B\mu_L^{(1)}}\frac{m_{\tilde{L}}^2}{\mu_d^2}~.
\label{sneu-vev}
\ea 
Here $m_{\tilde{L}}$ is the soft mass of the left-handed sleptons.
Thus, if $\mu_d^2 \gg m_{\tilde{L}}^2$, then $\langle
\tilde{\nu}_1\rangle \gg \langle H_d^0 \rangle$.

In the remainder of this paper, we will study the choice $R=R_1$ and
the case $\langle \tilde{\nu}_1\rangle \gg \langle H_d^0 \rangle$ in
detail as it gives rise to rather novel and interesting phenomenology.
We will discuss the phenomenology of the $R=R_2$ case in
Section~\ref{b-r} very briefly, since that case has already been
considered in~\cite{Brust:2011tb}.  For the case we are interested in,
the superpotential in (\ref{W-KPW}), (\ref{W-adj}) and the soft terms
in (\ref{soft-R}), (\ref{soft-new}) can be simplified since the large
$\mu_d$ term allows us to integrate out the fields $R_u$ and $H_d$,
leading to the following:
\ba
W &=& \mu_u H_u R_d +  \lambda^{S}_u  S H_{u} R_{d} +\lambda^{T}_u H_{u} T  R_{d} + y^{u}_{ij} H_{u} Q_{i} U^{c}_{j}  +   \lambda_{ijk} L_{i} L_{j} E^{c}_{k} +\lambda'_{ijk} L_{i} Q_{j} D^{c}_{k}~, 
\label{W-final}
\\ [0.5em]
L_{soft} &=& \sum_{i} m^{2}_{i} \Phi_{i}^{\dagger} \Phi_{i} 
+ \left[t_{S} S + \frac{1}{2} b_{S} S^{2}  + \frac{1}{2} b_{T} T^{2} + \frac{1}{2} b_{O} O^{2} + B\mu_L^{(i)}\, H_u L_i
\right.
\nonumber\\
&& \hspace{2.6cm}
\mbox{} + \left. \rule{0mm}{6mm}
 \frac{1}{3} A_{S} S^{3}+A^{\lambda}_T S T^2 + A^{\lambda}_O S O^2 + A_{S}^{(i)} S H_{u} L_{i} + A_{T}^{(i)} H_{u} T L_{i}+ {\rm h.c.} \right],
\label{Wsoft-final}
\vspace{-10pt}
\ea 
where the terms in the second line in $L_{soft}$ are assumed to be
suppressed relative to $M_{\rm SUSY}$ for simplicity, from the
arguments below (\ref{Wtrilinear}).  \emph{Note that in the $R = R_1$
scenario the down-type masses arise from the $L L E^c$ and $L Q D^c$
operators when the left-handed sneutrino gets a {\it vev}} (assuming
$\langle H^0_d \rangle \ll \langle \tilde{\nu}_1 \rangle$).

\subsection{$R$-breaking}
\label{r-break}

It is well-known that the vanishingly small value of the cosmological
constant breaks $R$-symmetry since it requires a non-zero value of the
superpotential in the vacuum, and the superpotential has non-zero
$R$-charge.  Since the gravitino mass $m_{3/2} \sim \langle W\rangle$
(in Planck units), this implies that $m_{3/2}$ is the order parameter
of $R$-breaking.  As mentioned in the Introduction, in this work we
imagine a setup in which $m_{3/2}$ is much smaller than the TeV scale.
Hence, the effects of $R$-breaking are also small.

The breaking of $R$-symmetry will eventually be transmitted to the
visible sector.  This can essentially happen in two ways.  A simple
possibility is that $R$-breaking is mediated to the visible sector by
generic Planck suppressed operators, which we denote as ``generic
gravity mediation".  This is a natural possibility since gravity is
expected to violate all global symmetries in general.  However,
another possibility is that these generic Planck suppressed operators
respect the $R$-symmetry (at least to a very good approximation) due
to it being an accidental symmetry of the visible sector, see
\cite{Kribs:2010md} for an example.  In this case $R$ breaking will
generically be communicated to the visible sector via anomaly
mediation.

In fact, the source of $R$-breaking can be connected to observable
physics in an interesting way.  For example, for $R=R_1$ the breaking
of the $R$-symmetry will give rise to neutrino masses
\cite{Frugiuele:2011mh,Bertuzzo:2012su}, while for $R=R_2$,
$R$-breaking will give rise to nucleon-antinucleon oscillations and
may also lead to proton decay in certain cases\cite{Brust:2011tb}.
Existing constraints from these observables then put an upper bound on
$m_{3/2}$ and therefore on the messenger scale, $M_*$, as well
\cite{Frugiuele:2011mh}.  In Appendix \ref{r-break}, we describe some
details of the sizes of $R$-breaking operators.  However, the collider
phenomenology is largely determined by the approximate $R$-symmetry,
and therefore we will often focus on the $R$-symmetric limit.  A more
thorough analysis of the full effects of $R$-breaking is left for the
future.

\section{Characteristic Features of the Model  with $R=R_1$} 
\label{charac}

In the $R$-symmetric limit, the superpotential and soft terms are
given by Eqs.~(\ref{W-final}) and (\ref{Wsoft-final}).  The terms
including the lepton (super) fields above are written in a general
basis.  In a general basis, all the sneutrino fields can develop
non-vanishing vacuum expectation values (\emph{vev}'s) since there are
$B\mu_L$ terms for all of them.  However, it is always possible to
choose a basis in which only one of the sneutrino fields gets a {\it
vev}\cite{Bertuzzo:2012su}.\footnote{The physics is of course basis
independent.} In this sense, there is a similarity with models of
$R$-parity violation (RPV) where the $H_d$ and $L_{i}$ fields mix, in
general, with each other, giving rise to mass eigenstates $L_{\alpha},
\alpha=1,...4$.  There is an important difference, however, since
$H_d$ has been integrated out, and the ``light down-type doublet",
$R_d$, has a different $R$-charge from the $L_i$.  Thus, there is only
a three-dimensional space in which the (sneutrino) fields can mix.  We
will follow the analysis in \cite{Barbier:2004ez} keeping this
difference in mind.

The  general basis can be related to the ``single {\it vev} basis" as:
\ba 
\label{singlevev}
L_i &=& \frac{v_{i}}{v_{(a)}}\,L_{(a)} + \sum_b\,e_{i\,b}\,L_b~.
\vspace{-10pt}
\ea 
Here the lepton of flavor $(a)$ is assumed to get a {\it vev}, while
$i,j,k$ run over all three generations and $b$ (and later $c$) runs
over only two generations (those which do not get a {\it vev}).  The
$e_{i\,b}$ are the matrix elements that relate the fields in the two
bases.  There is still the freedom to rotate $L_b$, and by choosing an
appropriate $e_{ib}$ one can go to a basis in which the charged Yukawa
couplings are diagonal.

Since the lepton Yukawa coupling is provided by the $\lambda_{ijk} L_i
L_j E^c_k$ operator, and $\lambda_{ijk}$ is antisymmetric in the first
two indices, gauge invariance prevents the lepton of flavor $(a)$ from
getting a mass from such operators.  Its mass can, nevertheless, be
generated from supersymmetry breaking (but $R$-preserving) operators.
For example, it could come from the following operators:
\ba 
\begin{array}{llll}
i)  &  
\displaystyle
y'_{(1)} \int \! d^4\theta \, \frac{X^{\dag}}{M_{\star}^2}\,H_u^{\dag}L_{(a)}\,E^c_{(a)}~, &   
ii) & 
\displaystyle
y'_{(2)} \int \! d^4\theta \, \frac{X^{\dag}X}{M_{\star}^2}\,H_u^{\dag}\,\frac{{\cal D}_{\alpha}\,L_{(a)}\,{\cal D}_{\alpha}\,E_{(a)}^c}{4\,\mu_d^2}~,
\label{emass} \\ [1em]
& \displaystyle
\Delta m^{(1)}_{L} ~=~ y'_{(1)}\frac{|F_X|}{M_{\star}^2}\,v_u \sim \left(\frac{M_{\rm SUSY}}{M_{\star}}\right)\,v_u~,  &  
& \displaystyle
\Delta m^{(2)}_{L} ~=~ y'_{(2)}\frac{|F_X|^2}{M_{\star}^2}\frac{v_u}{4\,\mu_d^2} \sim \left(\frac{M_{{\rm SUSY}}^2}{4\,\mu_d^2}\right)\,v_u~. 
\end{array}
\vspace{-15pt}
\ea 
We see that the first operator can provide viable lepton masses only
if the messenger scale $M_{\star}$ is low \cite{Davies:2011mp,
Frugiuele:2011mh}, but the second operator is generated after
integrating out the fields $R_u$ and $H_d$ with a large supersymmetric
mass term $\mu_d\,R_u H_d$ (see discussion around (\ref{sneu-vev}),
and appendix B in \cite{Kumar:2011np}), and can give a viable
contribution even if the messenger scale $M_{\star}$ is high.  Note
that both these operators are present for all lepton flavors in
general, so these will provide contributions to lepton masses \emph{in
addition} to those from the superpotential in
(\ref{W-final}).\footnote{They will also contain flavor off-diagonal
entries, presumably of order $m_e$.  In the lepton mass eigenbasis,
these will induce off-diagonal slepton masses, even if these are
diagonal (but not degenerate) in the gauge eigenbasis.  In this
$R$-symmetric framework, we expect the constraints from $\mu \to e
\gamma$ and $\mu-e$ conversion in nuclei to be satisfied since the mixing angles are 
of order $m_e/m_\mu$ and because we take the Dirac bino mass $M^D_1$ around
1~TeV (see section \ref{pheno}) ~\protect\cite{Fok:2010vk}.} Therefore, the smallness of the electron
mass makes it natural to take $(a)=1(e)$, and
$b,c=2\,(\mu),3\,(\tau)$, so that the electron gets its mass
\emph{solely} from supersymmetry breaking (but $R$-preserving)
operators in (\ref{emass})\cite{Bertuzzo:2012su}.  We will assume this
henceforth.

In the ``single-vev" {\it and} ``mass-eigenstate" basis, the
superpotential is given by:
\ba 
\label{convbasis}
W &=&  \mu_u H_u R_d+ \lambda^S_u S H_u R_d+\lambda^T_u  H_u T R_d+ W_{\mathrm{Yukawa}} + W_{\mathrm{Trilinear}}~,
\nonumber
\\ [0.7em]
W_{\mathrm{Yukawa}} &=& \sum_{b=2,3}\,y^{(e)}_{b} \hat{L}_{(a)} \hat{L}_b \hat{E}^c_b+ \sum_{i=1,2,3}\,y^{(d)}_{i} \hat{L}_{(a)} \hat{Q}_i \hat{D}_i^c~,
\\
W_{\mathrm{Trilinear}} &=& \sum_{i=1,2,3}\,\lambda_{2 3 i} \hat{L}_2 \hat{L}_3 \hat{E}^c_i+ \sum_{i,j=1,2,3; b=2,3}{ \lambda'_{b i j} \hat{L}_b \hat{Q}_i \hat{D}_j^c}~.
\nonumber
\ea
where the hat denotes that the lepton and quark fields are in the
``mass-eigenstate" basis.  Note that the first two indices in the
trilinear term $LLE^c$ in $W_{\rm{Trilinear}}$ in
Eq.~(\ref{convbasis}) are fixed to be (2) and (3) since $(a)=(1)$, and
since the coupling is antisymmetric in the first two indices.  There
is, however, no such antisymmetry for the first two indices in the
$LQD^c$ term in $W_{\rm{Trilinear}}$.  Overall, these terms have a
rather different flavor structure compared to analogous trilinear RPV
couplings in RPV models \cite{Barbier:2004ez}.

To contrast some other important features of our model against RPV
models considered in the literature, it is instructive to look at
properties of standard RPV models with both bilinear and trilinear RPV
operators, where we use the established results summarized in
\cite{Barbier:2004ez}.  For example, the presence of the fermionic and
scalar bilinear RPV operators in such models is associated with mixing
between neutrinos and neutralinos (or charged leptons and charginos).
In particular, the superpotential bilinear $\mu'_i\,\tilde{H}_u L_i$
generates neutrino masses at tree-level proportional to $\tan^2\xi$
through such mixings, where the angle $\xi$ parametrizes the physical
higgsino-lepton mixing in the fermion sector (which cannot be rotated
away) in a basis-independent way.  This leads to a very stringent
bound, $\sin \xi \lesssim 3\times10^{-6}\sqrt{1+\tan^2\beta}$.  Hence,
in the basis with a single $\mu$-term (i.e.~$\hat{\mu} H_u H_d$, but
no $\mu'_i H_u L_i$), the sneutrino {\it vevs} are forced to be
extremely small due to the upper bound on neutrino masses.
Furthermore, the presence of the scalar bilinear RPV operator
$B\mu_L^{(i)} H_u L_i$ in ${\cal L}_{soft}$ and the trilinear RPV
operators $\lambda \, LLE^c, \,\lambda' LQD^c$ in the superpotential,
give rise to one-loop contributions to neutrino masses proportional to
i) $g^2\,(B\mu_L^{(i)})^2$, ii) $g\lambda'$ (or $g\lambda$), iii)
$\lambda'^2$ or ($\lambda^2$), \footnote{Here $g$ schematically
denotes any of the two electroweak gauge couplings.} which put
stringent bounds on many $\lambda,\lambda'$ couplings (such as
$\lambda_{i33}$ and $\lambda'_{i33}$, $i=1,2,3$) as well as the size
of $B\mu_L^{(i)}$ (this can also be interpreted as putting a stringent
bound on the sneutrino {\it vev}'s through the basis-independent
Higgs-slepton mixing angle $\sin \zeta$, even in the absence of the
superpotential bilinear RPV operator).

In the model under consideration, the operators $\lambda\,LLE^c$ and
$\lambda' LQD^c$ in the superpotential, and the operator $B\mu_L^{(i)}
H_u L_i$ in ${\cal L}_{soft}$, preserve a lepton number (which is
identified with the $R$-symmetry $R=R_1$), unlike that in standard RPV
models above.  Hence these terms cannot generate Majorana neutrino
masses which violate the lepton number ($R=R_1$) either at tree-level
or loop-level, as long as the $R$-symmetry is conserved.  This further
implies that the sneutrino {\it vev} (induced by the $B\mu_L^{(i)}$
term) can be significant and can play the role of a Higgs field.  Also
note that there is no $\mu'_i \tilde{H}_u L_i$ term in the Lagrangian,
implying that the basis in which the Yukawa couplings in
$W_{\mathrm{Yukawa}}$ are diagonal is the same as the mass-eigenstate
basis of the charged leptons, unlike in RPV models
\cite{Barbier:2004ez}.

Finally, from above we see that the bounds on neutrino masses are only
relevant when $R$-breaking effects are taken into account and are,
hence, proportional to $m_{3/2}$ (see Section~\ref{r-break}).  Thus,
for given values of the sneutrino {\it vev} and the $\lambda,\lambda'$
couplings (which are consistent with other constraints, see
Section~\ref{sec:ewp}), the bounds from neutrino masses only provide a
bound on $m_{3/2}$.

\section{The Fermionic Electroweak Sector}
\label{fermionic-ew}

Since in our framework lepton number is identified with an
$R$-symmetry ($R=R_1$), the neutralinos and neutrinos on the one hand,
and charginos and charged leptons on the other, share the same quantum
numbers (in particular, their $R$-charge).  They can, therefore, mix
after electroweak symmetry breaking.

\subsection{Charginos $\&$ Charged Leptons}
\label{sec:charginos}

The charginos and charged leptons will mix after
electroweak symmetry breaking in general.  However, from
(\ref{convbasis}), it is clear that only the charged lepton of flavor
$(a)$ (the electron since $(a)=1$) will mix with the charginos since
only that flavor gets a \textit{vev}.  One has four \emph{Dirac}
charginos (one of which is the electron), which can be further split
according to their electric and $R$-charges.  Then, one can form two
groups of 2-component fields - one with $R=+Q$, and one with $R=-Q$.
This implies that the chargino mass matrix can be written as:
\ba 
\vspace{-10pt}
\mathcal{L}_{C} &=& \left((\tilde{\chi}^{++})^T \,(\tilde{\chi}^{+-})^T\right)
                                      \begin{pmatrix} 
                                      {\bf M}_C^{(+)} & 0 \cr 
                                      0 & {\bf M}_C^{(-)}\cr 
                                      \end{pmatrix}\, 
\begin{pmatrix}\tilde{\chi}^{--}\cr \tilde{\chi}^{-+}\end{pmatrix}~,
\\ [0.5em]
\tilde{\chi}^{++}&=&(\tilde w^+,  e_R^c)~,
\hspace{5mm}
\tilde{\chi}^{--} ~=~ (\tilde{T}_d^-, e_L^-)~,
\hspace{10.9mm}
(\textrm{for } R~= +\,Q)~
\nonumber \\ [0.3em]
\tilde{\chi}^{-+}&=&( \tilde w^- \tilde R_d^{-})~, 
\hspace{5.5mm}
\tilde{\chi}^{+-} ~=~ (\tilde{T}_u^+,\tilde{H}_u^{+})~,
\hspace{9mm}
(\textrm{for } R~= -\,Q)~
\nonumber
\ea 
where
\ba 
{\bf M}_C^{(+)} &=& 
\begin{pmatrix}
  M^D_{2} &  g v_{(a)}\cr
  0 & m_e \cr
\end{pmatrix}~,
\hspace{1cm}
{\bf M}_C^{(-)} = \begin{pmatrix}
  M^D_{2} &   \sqrt{2}\lambda^T_u v_u\cr
  g v_{u}  &  -\mu_u - \lambda^S_u v_s+\lambda^T_u v_T  \cr
\end{pmatrix}~.
\ea 
The notation $\tilde{\chi}^{+-}$, for instance, implies that the field
$\tilde{\chi}$ has electric charge +1 and $R$-charge -1.  Here
$M^D_{2}$ stands for the Dirac wino mass, $g$ is the $SU(2)$ gauge
coupling, $m_e$ is the electron mass, and $v_u$, $v_{(a)}$, $v_s$ $\&$
$v_T$ are the {\it vev}'s of $H_u^0$, $\tilde{\nu}_{(a)}$, $S$ and $T^0$
respectively.  The above matrices can be diagonalized by two pairs of
$2\times 2$ matrices - $\{{\bf V}^{+},{\bf U}^{+}\}$ for
$\left(\tilde{\chi}^{++}\, \tilde{\chi}^{--}\right)$, and $\{{\bf
V}^{-},{\bf U}^{-}\}$ for $\left(\tilde{\chi}^{+-}\,
\tilde{\chi}^{-+}\right)$, respectively, such that $({\bf V}^+)^{\dag}
{\bf M}_C^{(+)} {\bf U}^+$ and $({\bf V}^-)^{\dag} {\bf M}_C^{(-)}
{\bf U}^-$ are diagonal (and with positive eigenvalues).  The above
states are naturally arranged into four 4-component Dirac fields
$\tilde{X}_i^{++}=(\tilde{\chi}_i^{++}
\overline{\tilde{\chi}_i^{--}})$ and
$\tilde{X}_i^{+-}=(\tilde{\chi}_i^{+-}
\overline{\tilde{\chi}_i^{-+}})$, with $i=1,2$, whose charge
conjugates are denoted by $\tilde{X}_1^{--}$ and $\tilde{X}_i^{-+}$
respectively.  In this notation, $e \equiv \tilde{X}_1^{--}$
corresponds to the physical (Dirac) electron field.

\subsection{Neutralinos $\&$ Neutrinos}
\label{sec:neutral}

Similar to the charged fermion sector, the neutralinos and neutrinos in the
neutral fermion sector will mix after electroweak symmetry breaking.  For
simplicity, we consider only one neutrino generation since adding the
other neutrinos does not change the qualitative collider picture (see
\cite{Bertuzzo:2012su} for a thorough discussion of neutrino masses
and mixings).  Also, since Majorana neutrino masses violate lepton
number (hence $R$ symmetry in our case), they are massless in the
$R$-symmetric limit.\footnote{We are assuming that there are no
right-handed neutrinos with $R$-charges such as to allow writing down
Dirac neutrino mass terms consistent with the $R$-symmetry.} It is
convenient to write the matrix in the ``Dirac" basis by grouping the
fields with $R$-charges +1 and -1 separately.

Then, similar to the charginos, the mass terms for the neutralinos can
be written as ${\cal L}_N = (\tilde{\chi}^{0+})^T \, {\bf M}_N
\,\tilde{\chi}^{0-}$ where $ \tilde{\chi}^{0+}=( \tilde b^0,
\tilde{w}^0, \tilde R_d^{0})$ and $\tilde{\chi}^{0-}=( \tilde
S,\tilde{T}^0, \tilde H_u^{0}, \nu_e)$.  The notation
$\tilde{\chi}^{0+}$ implies that the field $\tilde{\chi}$ has
vanishing electric charge and $R$-charge +1.  The mass matrix $M_N$ is
given by:
\ba
\vspace{-10pt}
{\bf M}_N = \begin{pmatrix}
 M^{D}_1 & 0 & \frac{g' v_u}{\sqrt{2}} & -\frac{g' v_{(a)}}{\sqrt{2}}\cr
0 & M^D_2  &   -\frac{g v_u}{\sqrt{2}} & \frac{g v_{(a)}}{\sqrt{2}} \cr
 \lambda^S_u\,v_u & \lambda^T_u\,v_u  & \mu_u + \lambda^S_u\, v_s + \lambda^T_u\,v_T & 0 \cr
 \end{pmatrix} ~.
\ea
After diagonalizing the above mass matrix by unitary transformations
$\bf{V^N}$ and $\bf{U^N}$, as in the chargino case above, one obtains
three \emph{Dirac} mass eigenstates $\tilde{X}_i^{0+}\equiv
(\tilde{\chi}_i^{0+}\,\overline{\tilde{\chi}_i^{0-}})$, with
$i=1,2,3$, and one massless \emph{Weyl} neutralino
$\tilde{\chi}_4^{0-}$ that necessarily remains massless, which is
identified with the massless neutrino eigenstate, and is given in
general by:
\ba
\tilde{\chi}_4^{0-} = U^{N}_{4\tilde{s}}\,\tilde{S} +
U^{N}_{4\tilde{t}}\,\tilde{T}^0 + U^{N}_{4u}\,\tilde{H}_u^0+
U^{N}_{4\nu}\,\nu_e~.
\ea 
By a slight abuse of notation, sometimes we
will also refer to $\tilde{\chi}_4^{0-}$ as $``\nu_e"$, where it will
always refer to the mass eigenstate, and should not be confused with
the original gauge eigenstate.

\section{Indirect Constraints}
\label{sec:ewp}

The identification of the lepton numbers of SM fermions with (the
negative of) their $R$-charges is subject to various ``indirect"
constraints.  As explained in Section~\ref{charac}, the sneutrino {\it
vev} can be significant in these models since there are no bounds from
neutrino masses. Thus, the most stringent constraints in these
models arise from electroweak precision measurements and from flavor
violating processes \cite{Frugiuele:2011mh,Bertuzzo:2012su}.  We will
see that these can be satisfied in a large range of parameter space
and for reasonable flavor-off diagonal couplings.  Many of the
constraints are similar to those studied in \cite{Frugiuele:2011mh}.
It is convenient to divide the constraints into two categories:
\begin{itemize}
    
\item Constraints on the sneutrino vev $v_{(a)}$, or equivalently
$\tan \beta \equiv \frac{v_u}{v_{(a)}}$.

\item Constraints on the $\lambda$ and $\lambda'$ couplings.
\end{itemize}
%

\subsection{Constraints on the sneutrino vev $v_{(a)}$ ($\tan \beta$)}

\begin{wrapfigure}[10]{r}{0.48\textwidth}
\vspace{-22pt}
\centering
\includegraphics[width=0.3\textwidth]{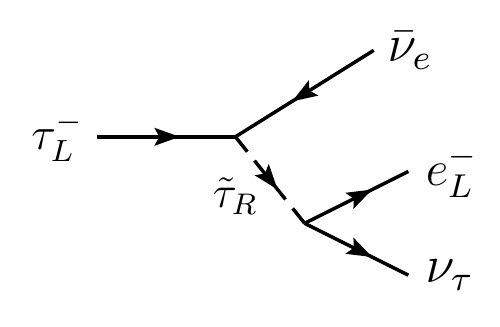}
\vspace{-10pt}
\caption{\footnotesize{Contribution to the decay width $\tau_L^-
\rightarrow e_L^-\overline{\nu}_e\,\nu_{\tau}$, where both the
interaction vertices correspond to the $\tau$ Yukawa coupling
$y_{\tau}\equiv \lambda_{133}$. The arrows indicate $R$-number flow.}}
\label{rtau}
\vspace{-20pt}
\end{wrapfigure}

First, the mixing between the charged leptons and charginos gives rise
to a deviation in the couplings of the $Z$ to charged leptons in
general, which is constrained by electroweak precision measurements.
As shown in the previous section, since only one lepton (of flavor
(a)=1) mixes with the charginos, the mixing is identical to that in
\cite{Frugiuele:2011mh}, giving rise to the following deviation in the
vector and axial-vector coupling to the $Z$ from those in the SM:
\ba
\label{Zcoupling}  
\vspace{-10pt} 
\delta\,g^i_V &=& \delta\,g^i_A ~=~ -\frac{\sin^2\phi}{2}~,
\\ 
\sin\phi &=&- \frac{\left(m_e^2+g^2\,v_{(a)}^2-(M^D_2)^2\right)+
\sqrt{\left[m_e^2+g^2\,v_{(a)}^2+(M^D_2)^2 \right]^2-4\,m_e^2 (M^D_2)^2}}{2\,g\,v_{(a)} M^D_2}~. 
\nonumber
\ea 
From the measured value of the coupling $g_A^{e}=-0.50111\pm 0.00035$
\cite{Loinaz:2004qc}, one gets an upper bound on
$\left(\frac{v_{(a)}}{M^D_{2}}\right)$.  For example, for $M^D_{2}$=
1500 GeV (300 GeV), the current data puts an upper bound $v_{(a)}
\lesssim 61$ GeV (12 GeV) at 1$\sigma$.  In this work, we assume that
the (Dirac) gauginos are heavier than the scalars.  As a benchmark, we
take $M^D_2 \simeq 1.5$ TeV henceforth, implying that $v_{(a)}^{max}
\simeq 60$ GeV. The fact that only the charged lepton of flavor $(a)$
($e$ in our case) mixes with the charginos also gives rise to
constraints from charged current universality.  However, the bounds
from these are not as strong as those derived from the $Z$-coupling
above (see \cite{Frugiuele:2011mh}).  A lower bound on $v_{(a)}$,
however, arises from leptonic Yukawa couplings, $y_{\tau}$ in
particular.  This is because the leptonic Yukawa couplings arise from
the $LLE^c$ operator, and hence are part of the $\lambda$ couplings [see
Eq.~(\ref{W-final})].  Therefore, these give rise to extra tree-level
contributions to electroweak observables similar to those in
traditional RPV models (for a review of constraints in RPV models, see
\cite{Barbier:2004ez}).  One finds that the dominant constraint arises
from the $\tau$ Yukawa coupling ($y_{\tau}\equiv y_3^{(e)}\equiv
\lambda_{133}$) contributing to the ratio $R_{\tau} \equiv \Gamma(\tau
\rightarrow e\,\bar{\nu}_e\,\nu_{\tau}) / \Gamma(\tau \rightarrow
\mu\,\bar{\nu}_{\mu}\,\nu_{\tau})$ \cite{Frugiuele:2011mh}, as shown
in Fig.~\ref{rtau} ($R_\tau$ is normalized to the dominant $\tau$
decay due to $W$ exchange).  This gives:
\ba
\label{ytau} 
y_{\tau} &<& 0.07\,\left(\frac{m_{\tilde{\tau}_{R}}}{100\,{\rm GeV}}\right)~, 
\ea 
which puts a lower bound $v_{(a)} \gtrsim 8$ GeV (2.5 GeV) for
$m_{\tilde{\tau}_{R}}= 300$ GeV (1 TeV).  Since in our
framework, one of the sneutrinos behaves as a Higgs field which is
expected to be around the electroweak scale, it is natural to expect
that the masses of all sleptons are around the electroweak scale,
i.e.~few hundred GeV (see more discussion on this in
Section~\ref{pheno}).  We, therefore, take $v_{(a)}^{min} \simeq 10$
GeV. Combining with the upper bound on $v_{(a)}$ from above, we obtain
the range:
\ba 
10~{\rm GeV} &\lesssim& v_{(a)} ~\lesssim~ 60~{\rm GeV}~, 
\hspace{1cm}
\nonumber\\
{\rm or}
\hspace{1cm}
17.4 &\gtrsim& \tan\,\beta ~\gtrsim~ 2.7~,
\label{tanbrange}
\ea 
where $\tan\beta = v_u/v_{(a)}$, which we use in our subsequent
analysis.  Thus, we see that the sneutrino {\it vev} can be much
larger than in standard (bilinear) RPV models.  This is one of the
most distinctive features of the model, and it also plays a crucial
role in LHC phenomenology as we will see.  In Section~\ref{pheno}, we
will discuss how the existence of such a large sneutrino {\it vev}
could be inferred at the LHC.

\subsection{Constraints on $\lambda$ and $\lambda'$ couplings}
\label{llp-const}

As explained at the end of Section~\ref{charac}, in our model the
bounds from neutrino masses can be interpreted as a bound on the
gravitino mass (since it is the order parameter of $R$-breaking),
implying that the bounds on $\lambda,\lambda'$ couplings only arise
from other observables, such as flavor-violating processes.  We will
see that this has an important effect on the bounds on the
$\lambda_{i33}$ and $\lambda'_{i33}$ ($i=1,2,3$) couplings in
particular, because in standard RPV models the most stringent bounds
($\lambda_{i33}, \lambda'_{i33} \lesssim 10^{-3}$) on these arise from
neutrino masses.\footnote{In standard RPV models, the contributions
from $\lambda, \lambda'$ couplings are proportional to the quark and
lepton masses, hence the bounds are rather tight for the third
generation (s)quarks: $\lambda_{i33}$ and $\lambda'_{i33}$, for
$i=1,2,3$.} On the other hand, the upper bounds on $\lambda_{i33}$ and
$\lambda'_{i33}$ in our model arise from flavor-violating processes
and can be rather mild as we will see.

Constraints from flavor-violating processes in the lepton and hadron
sector provide bounds on the $\lambda$ and $\lambda'$ couplings.
Although there do exist bounds on single couplings, they are typically
weak, and most of the stringent bounds arise from the products of two
couplings.  These can be classified into three categories, those
constraining i) $\lambda\,\lambda$ couplings, ii) $\lambda\,\lambda'$
couplings, and iii) $\lambda'\,\lambda'$ couplings.  We take our
results from the general analysis of constraints arising from RPV
operators in \cite{Barbier:2004ez, Saha:2002kt, Dreiner:2006gu,
Dreiner:2012mx}.  However, as mentioned above, in our framework the
lepton and down-type Yukawa couplings are part of the $\lambda$ and
$\lambda'$ couplings respectively, and the results of
\cite{Barbier:2004ez, Saha:2002kt, Dreiner:2006gu, Dreiner:2012mx}
must be interpreted accordingly.  In particular, using the notation in
(\ref{convbasis}), $\lambda_{122}\equiv y_2^{(e)}$ and
$\lambda_{133}\equiv y_3^{(e)}$, while $\lambda'_{111}\equiv
y_1^{(d)}$, $\lambda'_{122}\equiv y_2^{(d)}$, and
$\lambda'_{133}\equiv y_3^{(d)}$.  The remaining $\lambda,\,\lambda'$
couplings are included in $W_{{\rm Trilinear}}$ in (\ref{convbasis}).
Among these non-Yukawa trilinear couplings, there are only three
independent couplings of the $\lambda$-type\footnote{Note that the
first two indices in the $ \lambda$ couplings are anti-symmetric.} --
$\lambda_{23i}$ ($i=1,2,3$), and two $3\times 3$ matrices of the
$\lambda'$-type -- $\lambda'_{2jk}$ and $\lambda'_{3jk}$ with
independent entries.  Thus, not only the number of independent
couplings is greatly reduced, but also the implications of these
bounds on the parameter space of the model are very different compared
to those within a standard RPV scenario.

\begin{table}[h!]
\begin{center}
\begin{tabular}{|c|c|c|}
\hline
\rule{0mm}{5mm}
{\bf Coupling(s)} & {\bf Upper Bound}  & {\bf Process} \\
[0.3em]
\hline
\rule{0mm}{5mm}
$|\lambda_{23k}|$;\,\footnotesize{k=1,2,3} & $0.07\,\tilde{e}_{kR}$ &$[R_{\tau},\,R_{\tau\mu}]$ \cite{Barbier:2004ez, Saha:2002kt}\\
[0.3em]
\hline
$|\lambda_{231}\,y_3^{(e)}|$ & $5.2\times 10^{-4}\,[\tilde{\nu}_{L_2}]^2$ &  \multirow{2}{*}{$[\tau \rightarrow e\,e\,\bar{\mu}]$ \cite{Dreiner:2006gu}}  \\ 
[0.3em]
or,\;$|\lambda_{231}|$ &  $0.052\,\cos \beta\,[\tilde{\nu}_{L_2}]^2$ &   \\
[0.3em]
\hline
$|\lambda_{232}\,y_3^{(e)}|$ & $7.0\times 10^{-4}\,[\tilde{\nu}_{L_2}]^2$ &   \multirow{2}{*}{$[\tau \rightarrow \mu\,e\,\bar{\mu}]$ \cite{Dreiner:2006gu}} \\
[0.3em]
or,\;$|\lambda_{232}|$ &  $0.070\,\cos \beta\,[\tilde{\nu}_{L_2}]^2$ &\\
[0.3em]
\hline
$|\lambda_{233}\,y_3^{(e)}|$ & $ 2.2\times 10^{-4}$ &  \multirow{2}{*}{ [$\tau\rightarrow e P^0 /\, \mu-e$ {\rm in nuclei}] \cite{Dreiner:2012mx}} \\
[0.3em]
or,\;$|\lambda_{233}|$ &  $0.022\,\cos \beta$ &  \\
[0.3em]
\hline
\end{tabular}
\end{center}
\vspace{-10pt}
\caption{\footnotesize{Some upper bounds on $\lambda$ couplings from
various flavor-violating processes.  Here $[\tilde{\nu}_{L_2}]^2 \equiv
(\frac{m_{\tilde{\nu}_{L_2}}}{100\,{\rm GeV}})^2$.  The notation for
the processes is the same as in the corresponding references.}}
\label{lambda-bounds}
\end{table}
\begin{table}[h!]
\begin{center}
\begin{tabular}{|c|c|c|}
\hline
\rule{0mm}{5mm}
{\bf Coupling(s)} & {\bf Upper Bound(s)}  & {\bf Process} \\
[0.3em]
\hline
$|\lambda'_{211}\,y_2^{(e)}|$ & $2.1\times 10^{-8}\,[\tilde{\nu}_{L_2}]^2$ & \multirow{2}{*}{[$\mu-e$ in nuclei]\cite{Barbier:2004ez, Saha:2002kt} }  \\
[0.3em]
or,\;$|\lambda'_{211}|$ &  $3.5\times10^{-5}\,\cos \beta\,[\tilde{\nu}_{L_2}]^2$ &  \\
[0.3em]
\hline
$|\lambda'_{212}\,y_2^{(e)}|$ & $6.0\times 10^{-9}\,[\tilde{\nu}_{L_2}]^2$ & \multirow{2}{*}{ $[K_L^0\rightarrow \mu\bar{e}/e\bar{\mu}]$ \cite{Dreiner:2006gu}}  \\
[0.3em]
or,\;$|\lambda'_{212}|$ &  $1.01\times 10^{-5}\,\cos \beta\,[\tilde{\nu}_{L_2}]^2$ & \\
[0.3em]
\hline
$|\lambda'_{213}\,y_2^{(e)}|$ & $ 1.3\times 10^{-5}\,[\tilde{\nu}_{L_2}]^2$ &  \multirow{2}{*}{ [$B_d^0\rightarrow e\bar{\mu}$] \cite{Dreiner:2006gu}} \\
[0.3em]
or,\;$|\lambda'_{213}|$ &  $0.022\,\cos \beta\,[\tilde{\nu}_{L_2}]^2$ & \\
[0.3em]
\hline
$|\lambda'_{221}\,y_2^{(e)}|$ & $6.\times 10^{-9}\,[\tilde{\nu}_{L_2}]^2$ &   \multirow{2}{*}{$[K_L^0\rightarrow \mu\bar{e}/e\bar{\mu}]$\cite{Dreiner:2006gu}}  \\
[0.3em]
or,\;$|\lambda'_{221}|$ &  $1.01\times10^{-5}\,\cos \beta\,[\tilde{\nu}_{L_2}]^2$ &\\
[0.3em]
\hline
$|\lambda'_{222}\,y_2^{(d)}|$ & $1.0\times 10^{-5}$ &  \multirow{2}{*}{ [$\tau\rightarrow e P^0/\,\mu-e$ {\rm in nuclei}] \cite{Dreiner:2012mx} } \\
[0.3em]
or,\;$|\lambda'_{222}|$ &  $0.032\,\cos \beta$ & \\
[0.3em]
\hline
$|\lambda'_{223}\,y_2^{(e)}|$ & $ 7.6\times 10^{-5}\,[\tilde{\nu}_{L_2}]^2$ &     \multirow{2}{*}{[$B_s^0\rightarrow e\bar{\mu}$]  \cite{Dreiner:2006gu}} \\
[0.3em]
or,\;$|\lambda'_{223}|$ &  $0.128\,\cos \beta\,[\tilde{\nu}_{L_2}]^2$ &   \\
[0.3em]
\hline
$|\lambda'_{231}\,y_2^{(e)}|;|\lambda'_{231}\,y_3^{(d)}|$ & $1.3\times 10^{-5}\,[\tilde{\nu}_{L_2}]^2;\;1.6\times 10^{-3}\,[\tilde{u}_{L_3}]^2$ &       \multirow{2}{*}{[$B_d^0\rightarrow \mu\bar{e}$]  \cite{Dreiner:2006gu}} \\
[0.3em]
or,\;$|\lambda'_{231}|$ &  $0.022\,\cos \beta\,[\tilde{\nu}_{L_2}]^2;\;0.099\,\cos \beta\,[\tilde{u}_{L_3}]^2$ & \\
[0.3em]
\hline
$|\lambda'_{232}\,y_2^{(e)}|;|\lambda'_{232}\,y_3^{(d)}|$ & $7.6\times 10^{-5}\,[\tilde{\nu}_{L_2}]^2;\;2.7\times 10^{-4}\,[\tilde{u}_{L_3}]^2$ &     \multirow{2}{*}{ [$B_s^0\rightarrow \mu\bar{e}$]  \cite{Dreiner:2006gu};\;[$b\rightarrow s \mu \bar{e}$] \cite{Saha:2002kt}}  \\
[0.3em]
or,\;$|\lambda'_{232}|$ &  $0.128\,\cos \beta\,[\tilde{\nu}_{L_2}]^2;\;0.016\,\cos \beta\,[\tilde{u}_{L_3}]^2$ &\\
[0.3em]
\hline
$|\lambda'_{233}\,y_3^{(d)}|$ & $ 1.1\times 10^{-5}$ &  \multirow{2}{*}{[$\tau\rightarrow e P^0/\,\mu-e$ {\rm in nuclei}] \cite{Dreiner:2012mx}} \\
[0.3em]
or,\;$|\lambda'_{233}|$ &  $6.8\times 10^{-3}\,\cos \beta$ &   \\
[0.3em]
\hline
\end{tabular}
\end{center}
\vspace{-10pt}
\caption{\footnotesize{Some upper bounds on
$\lambda'_{2jk}$ ($j,k=1,2,3$) couplings from various flavor-violating
processes.}}
\label{lambdap2-bounds}
\end{table}

Starting with the $\lambda$ couplings, we find the bounds in Table
\ref{lambda-bounds}.  We have only listed single coupling bounds, and
those product ($\lambda\,\lambda$) bounds in which one of the
couplings is a Yukawa coupling.  This is because the Yukawa couplings
are known up to $\tan\,\beta$, hence they provide the most robust
bounds (however, we do consider the full set of constraints in our
analysis).  The $\lambda'$ couplings are more numerous.  Again, upper
bounds exist on various products of the form $\lambda\,\lambda'$ and
$\lambda'\,\lambda'$, and are listed in \cite{Barbier:2004ez,
Saha:2002kt, Dreiner:2006gu, Dreiner:2012mx}.  As before, we describe
the bounds on those products, one of which is a Yukawa coupling
(either of the leptonic type or the down-type quark type).  For some
couplings, bounds exist from more than one experiment.  We list the
dominant bound in such cases, unless the bounds are comparable in
which case we list all of them.  The bounds for $\lambda'_{2jk}$ and
$\lambda'_{3jk}$ are listed in Tables \ref{lambdap2-bounds} and
\ref{lambdap3-bounds} respectively.  As explained earlier, the bounds
on $\lambda_{i33}$ and $\lambda'_{i33}$ ($i=1,2,3$) are relaxed
relative to those in standard RPV models.  {\it In particular, the
bound on $\lambda'_{333}$ is very mild from Table
\ref{lambdap3-bounds}; it can be comparable to the electroweak gauge
couplings}.  This can have an important effect on collider
phenomenology related to the third generation (see
Section~\ref{pheno}).
\begin{table}[h!]
\begin{center}
\begin{tabular}{|c|c|c|}
\hline
\rule{0mm}{5mm}
{\bf Coupling(s)} & {\bf Upper Bound(s)}  & {\bf Process} \\
[0.3em]
\hline
$|\lambda'_{311}\,y_3^{(e)}|$ & $8.5\times 10^{-5}\,[\tilde{\nu}_{L_3}]^2$ & \multirow{2}{*}{ [$\tau\rightarrow e \eta$]\cite{Dreiner:2006gu}}  \\
[0.3em]
or,\;$|\lambda'_{311}|$ &  $8.47\times 10^{-3}\,\cos \beta\,[\tilde{\nu}_{L_3}]^2$ &  \\
[0.3em]
\hline
$|\lambda'_{312}\,y_3^{(e)}|$ & $9.7\times 10^{-4}\,[\tilde{\nu}_{L_3}]^2$ &   \multirow{2}{*}{ $[\tau\rightarrow e\,K_s]$ \cite{Dreiner:2006gu}}  \\
[0.3em]
or,\;$|\lambda'_{312}|$ &  $0.097\,\cos \beta\,[\tilde{\nu}_{L_3}]^2$ & \\
[0.3em]
\hline
$|\lambda'_{313}\,y_3^{(e)}|$ & $ 3.7\times 10^{-4}\,[\tilde{\nu}_{L_3}]^2$ &    \multirow{2}{*}{ [$B_d^0\rightarrow e\bar{\tau}$] \cite{Dreiner:2006gu}} \\
[0.3em]
or,\;$|\lambda'_{313}|$ &  $0.037\,\cos \beta\,[\tilde{\nu}_{L_3}]^2$ &  \\
[0.3em]
\hline
$|\lambda'_{321}\,y_3^{(e)}|$ & $9.7\times 10^{-4}\,[\tilde{\nu}_{L_3}]^2$ &    \multirow{2}{*}{$[\tau \rightarrow e\,K_s]$\cite{Dreiner:2006gu}}   \\
[0.3em]
or,\;$|\lambda'_{321}|$ &  $0.097\,\cos \beta\,[\tilde{\nu}_{L_3}]^2$ &\\
[0.3em]
\hline
$|\lambda'_{322}\,y_3^{(e)}|$ & $4.6\times 10^{-4}\,[\tilde{\nu}_{L_3}]^2$ &    \multirow{2}{*}{ [$\tau\rightarrow e \eta$] \cite{Dreiner:2006gu} } \\
[0.3em]
or,\;$|\lambda'_{322}|$ &  $0.046\,\cos \beta\,[\tilde{\nu}_{L_3}]^2$ & \\
[0.3em]
\hline
$|\lambda'_{331}\,y_3^{(e)}|;|\lambda'_{331}\,y_3^{(d)}|$ & $3.7\times 10^{-4}\,[\tilde{\nu}_{L_3}]^2;\;2.7\times 10^{-3}\,[\tilde{u}_{L_3}]^2$ &    \multirow{2}{*}{  [$B_d^0 \rightarrow \tau\bar{e}$] \cite{Dreiner:2006gu}  } \\
[0.3em]
or,\;$|\lambda'_{331}|$ &  $0.037\,\cos \beta\,[\tilde{\nu}_{L_3}]^2;\;0.168\,\cos \beta\,[\tilde{u}_{L_3}]^2$ & \\
[0.3em]
\hline
$|\lambda'_{333}\,y_3^{(d)}|$ & $2.1\times 10^{-2}$ &  \multirow{2}{*}{ [$l_i\rightarrow 3\,l_j$] \cite{Dreiner:2012mx}}  \\
[0.3em]
or,\;$|\lambda'_{333}|$ &  $1.305\,\cos \beta$ &  \\
[0.3em]
\hline
\end{tabular}
\end{center}
\vspace{-10pt} 
\caption{\footnotesize{Some upper bounds on $\lambda'_{3jk}$
($j,k=1,2,3$) couplings from various flavor-violating processes.  We
do not show bounds on $\lambda'_{323}$ and $\lambda'_{332}$ as they do
not appear in product bounds where the other coupling is a Yukawa
coupling.}}
\label{lambdap3-bounds}
\end{table}

In order to get a better idea about the constraints on these
couplings, it is useful to understand what generic expectations we
have for the spectrum of the model.  This will be discussed more in
Section~\ref{pheno}; here we just make some brief remarks.  We imagine
a situation in which the (Dirac) gauginos are heavy and the scalars
and Higgsinos are relatively light.  For concreteness, we take
$\mu=200$ GeV, $m_{\tilde{L}}^2 \simeq m_{\tilde{\bar{E}}}^2 \simeq
(200-300\,{\rm GeV})^2, M^D_1 = 1000\,{\rm GeV}, M^D_2\simeq
1500\,{\rm GeV}$.  The masses of the squarks are determined from
current LHC constraints, which are studied in detail in the companion
paper \cite{Frugiuele:2012xx}.  It turns out that the bounds on masses
of the first two generation squarks are in the 600-700 GeV range,
while the bounds on the third generation squark masses are lower,
around 400 GeV.

With the above spectrum in mind, it is straightforward (although
tedious) to check that (almost) all of the remaining bounds on the
products of $\lambda,\lambda'$ couplings in \cite{Barbier:2004ez,
Saha:2002kt, Dreiner:2006gu, Dreiner:2012mx} can be satisfied if the
values of the $\lambda,\lambda'$ couplings are assumed to saturate the
bounds in Tables \ref{lambda-bounds}, \ref{lambdap2-bounds} and
\ref{lambdap3-bounds}.  The exception is the
following product bound:
\begin{table}[h!]
\begin{center}
\tabcolsep 5.8pt
\small
\begin{tabular}{|c|c|c|}
\hline
\rule{0mm}{5mm}
{\bf Coupling(s)} & {\bf Upper Bound}  & {\bf Process} \\
[0.3em]
\hline
	\multirow{2}{*}{$|\lambda_{231}\,\lambda'_{311}|$} & \multirow{2}{*}{$2.1\times 10^{-8}\,[\tilde{\nu}_{L_3}]^2$} & \multirow{2}{*}{[$\tau\rightarrow e\,P^0 / \mu-e $ in nuclei]\cite{Dreiner:2012mx}}\\
[0.8em]
\hline
\end{tabular}
\end{center}
\vspace{-20pt}
\end{table}

A simple choice, therefore, is to assume that the coupling
$\lambda'_{311}$ is negligible ($\simeq 0$), while all the other
couplings still saturate the bounds in Tables \ref{lambda-bounds},
\ref{lambdap2-bounds} and \ref{lambdap3-bounds}.  We will see in
Section~\ref{pheno} that this leads to rather interesting prospects
for various signals such as lepto-quark-like signals, and single
slepton/sneutrino production.  Finally, it is worth mentioning that it
is possible to make other (reasonable) ans\"atze about the flavor
dependence of these couplings.  In Appendix \ref{ansatz}, we discuss
another simple ansatz about the flavor dependence of these couplings
which allows a nice understanding of the relative magnitudes of the
various couplings and also satisfies existing constraints.  It also
has the advantage of \emph{further} reducing the number of independent
$\lambda$ and $\lambda'$ couplings.

\section{The Scalar Electroweak Sector}
\label{higgs}

In this section, we discuss some important features of the electroweak
scalar sector of the model, and compare and contrast it with the well
known case of the MSSM. This is of considerable importance, especially
after the recent discovery of a Higgs-like particle near 125 GeV
\cite{:2012gu, :2012gk}.  A more thorough
treatment of these issues, including the couplings of the Higgs-like
particles in the model, will be the subject of another work
\cite{Kumar:2012yy}.

From Eqs.~(\ref{W-final}) and (\ref{Wsoft-final}), the electroweak
scalar potential takes the form:
\ba
\label{potential}
&&V^{EW} = V_F^{EW} + V_D^{EW} + V^{EW}_{\rm soft} + V^{(1)}_{\rm loop}~,\\
V_F^{EW} &=& \sum_i \left| \frac{\partial W}{\partial \phi_i} \right|^2~,
\hspace{1cm}
V_D^{EW} = \frac{1}{2}\,\sum_{a=1}^3 (D_2^a)^2 + \frac{1}{2}\, D_Y^2~,
\\
V_{\rm soft}^{EW} &=& m_{H_u}^2 |H_u|^2 + m_{R_d}^2 |R_d|^2 + \sum_{i=1}^3 m_{\tilde{L}_i}^2 |\tilde{L}_i|^2 + m_{s}^2 |S|^2 + 
m_{T}^2 T^{a \dagger} T^{a} + t_s\,S + \frac{1}{2}b_{S} S^2 + \frac{1}{3}A_s\,S^3 + 
\nonumber \\
&& \mbox{} \frac{1}{2}b_{T} T^a T^{a} + B\mu_L^{(i)}\,H_u L_i+ A_T\,ST^2 + A_{S}^{(i)}\,S H_u L_i+ A_{T}^{(i)}\, H_u T L_i+{\rm h.c.}~,
\nonumber
\ea
while $V^{(1)}_{\rm loop}$ refers to the one-loop contribution to the
effective potential, which will be specified below.  Here $i$ runs
over all the electroweak scalar fields which could receive {\it vev}'s,
and $D_Y$ and $D_2^a$ are the hypercharge and $SU(2)_{L}$ $D$-terms,
respectively.  Compared to the MSSM case, the $D$-terms contain
additional pieces associated with the $SU(2)_{L}$ and $U(1)_{Y}$
adjoint fields:
\ba
\label{D}
D_2^a &=& g (H_u^{\dag}\tau^a H_u + R_d^{\dag}\tau^a R_d + \tilde{L}_i^{\dag} \tau^a \tilde{L}_i+ T^{\dag} \lambda^a T) + 
\sqrt{2}\,(M^D_{2}\,T^a + {\rm h.c.})~,\\ [0.3em]
D_Y &=& \frac{g'}{2} (H_u^{\dag} H_u - R_d^{\dag} R_d-\tilde{L}_i^{\dag} \tilde{L}_i)  + \sqrt{2}\,(M^D_{1} S + {\rm h.c.})~,
\nonumber
\ea
where $\tau^a$ and $\lambda^a$ are the two and three-dimensional
$SU(2)$ generators respectively.  Note that the $D$ terms above give
rise to new \textit{trilinear} couplings in the scalar potential.
Also, the masses of the real and imaginary parts of $S = S_{R} + i \,
S_{I}$ and $T = T_{R} + i \, T_{I}$ are split in
Eq.~(\ref{potential}).  For instance, if $M^D_{i}$, $b_{S}$ and
$b_{T}$ are real, then $m_{S_{R}}^2 = m_{s}^2 + b_{S} + 4 (M^D_{1})^2$
and $m_{S_{I}}^2 = m_{s}^2 - b_{S}$ while $m_{T_{R}}^2 = m_{T}^2 +
b_{T} + 4 (M^D_{2})^2$ and $m_{T_{I}}^2 = m_{T}^2 - b_{T}$.  For
simplicity, we assume that there are no CP-violating phases in the
potential.

In order to minimize the above potential, we point out some important
simplifications.  First, EW precision constraints on the
$\rho$-parameter require the triplet Higgs {\it vev}, $\langle
T^{3}\rangle \equiv v_{T}$, to be small ($\lesssim 3$ GeV
\cite{Amsler:2008zzb}), which is naturally achieved if the triplet
soft breaking mass $m_{T} \gtrsim$ TeV. Therefore, the effect of the
triplet on the minimization of the potential must be small, and
$v_{T}$ can be set to zero in the first approximation.  Second, since
the $R$-symmetry forbids the term $B\mu\,H_u R_d$, it is easy to see
that $\langle R_{d} \rangle = 0$ if $m_{R_{d}}^{2} > 0$, i.e.~there is
no spontaneous breaking of the $U(1)_{R}$ symmetry.  Also, because
$R_d$ has a different $R$-charge (= 2) than the rest of the
electroweak fields (= 0), the degrees of freedom in $R_{d}$ do not mix
with those in the other fields and decouple from the rest.

It is important to understand the similarities and differences in the
structure of the scalar potential relative to that in ``supersoft"
SUSY breaking studied in \cite{Fox:2002bu}.  Since the gauge sector of
the model is similar to that in \cite{Fox:2002bu}, the model shares
the good feature that unlike the MSSM, the usual logarithmic
divergence from the stop contributions to the Higgs mass-squared
parameter $m_{H_u}^2$ are cutoff by the Dirac gluino mass, leading to
only a \emph{finite} contribution.  Thus, in contrast to the MSSM, a
Dirac gluino mass in the multi-TeV range is consistent with
electroweak-naturalness.

On the other hand, the matter sector of the model is rather different
from that in \cite{Fox:2002bu}.  Indeed, in the latter case one gets a
vanishing $D$-term contribution to the Higgs quartic coupling at tree
level, which is obviously not a good starting point to obtain a Higgs
mass near 125 GeV. In our model, however, the above conclusion is
circumvented by the presence of soft (but not supersoft) operators
arising from $F$-terms of the $X$ spurion in (\ref{softops}) which
yield the soft parameters $\{m_s^2, m_T^2, b_S, b_T\}$, etc.  Also,
the presence of the superpotential couplings in (\ref{W-final})
proportional to $\lambda^S_u$ and $\lambda^T_u$ can give rise to new
$F$-term contributions to the Higgs quartic coupling at tree level if
$R_d$ gets a {\it vev}, as in \cite{Belanger:2009wf, Benakli:2011kz}.
However, since in our model $\langle R_d \rangle = 0$, this tree-level
$F$-term contribution is not present.  Nevertheless, the $\lambda^S_u$
and $\lambda^T_u$ couplings do provide important contributions to the
Higgs quartic coupling at loop level.  We will see in the next
subsection that this is very important in obtaining a CP-even mass
eigenstate with mass $\sim 125$ GeV.

With the above simplifications, it suffices to minimize the scalar
potential with respect to the neutral fields -
$\{H_u^0,\,\tilde{\nu}_{(a)},\,S_R\}$ to study electroweak symmetry
breaking (EWSB):\footnote{For numerical results we do a full analysis,
including all \textit{vev}'s, and based on the full Coleman-Weinberg
potential.}
\ba
0&\simeq&\textstyle{\mu^2+m_{H_u}^2-\left(\frac{g^2+g'^2+4 \delta\lambda_u}{4}\right) v^2 c_{2\beta}+\left(\frac{2\delta\lambda_u+\delta\lambda_3}{2}\right) v^2 c_{\beta}^2 +\sqrt{2} g' v_s M^D_1+ \lambda^S_u v_s (2\mu+\lambda^S_u v_s)+ t_{\beta}^{-1} B\mu_L^{(a)}},
\notag\\ 
0&\simeq& \textstyle{m_{\tilde{L}_{(a)}}^2+\frac{(g^2+g'^2-\delta\lambda_3 + 2\delta\lambda_{\nu})}{4} \, v^2 c_{2\beta}+\left(\frac{\delta\lambda_3+2\delta\lambda_{\nu}}{4}\right) v^2 - \sqrt{2}g' v_s M^D_1 + t_{\beta} B\mu_L^{(a)}}~, 
\\ 
0&\simeq& \textstyle{[m_{S_R}^2 + (\lambda^S_u)^2\,v^2 s_{\beta}^2] \,v_s -\frac{g'}{\sqrt{2}}M^D_1 v^2 c_{2\beta} + (t_S + \lambda^S_u\,\mu\,v^2 s_{\beta}^2)}~.
\notag
\label{ewsb-eq}
\ea
Here, $s_{\beta}$ stands for $\sin\beta$ and so on, and
$\{\delta\lambda_u, \delta\lambda_{\nu},\delta\lambda_3\}$ denote the
dominant radiative corrections to the quartic terms:
$\frac{1}{2}\delta\lambda_u\,(|H_u^0|^2)^2$,
$\frac{1}{2}\delta\lambda_{\nu}\,(|\tilde{\nu}_{(a)}|^2)^2$ and
$\frac{1}{2}\delta\lambda_3\,|H_u^0|^2|\tilde{\nu}_{(a)}|^2$
respectively.  In the limit where $\lambda^S_u$ is negligible, the
leading-logarithm contributions to these radiative corrections are
given by \cite{Belanger:2009wf}:
\ba 
\label{deltalambda}
\delta\lambda_u &\simeq& \frac{3\,y_t^4}{16\pi^2}\,\log\left(\frac{m_{\tilde{t}_1}\,m_{\tilde{t}_2}}{m_t^2}\right) +  \frac{5\,(\lambda^T_u)^4}{16\pi^2}\,\log\left(\frac{m_T^2}{v^2}\right)~,
\nonumber\\
\delta\lambda_{\nu} &\simeq& \frac{3\,y_b^4}{16\pi^2}\,\log\left(\frac{m_{\tilde{b}_1}\,m_{\tilde{b}_2}}{m_t^2}\right) +  \frac{5\,(\lambda^T_u)^4}{16\pi^2}\,\log\left(\frac{m_T^2}{v^2}\right)~,
\\
\delta\lambda_3 &\simeq&   \frac{5\,(\lambda^T_u)^4}{32\pi^2}\,\log\left(\frac{m_T^2}{v^2}\right)~,
\nonumber
\ea 
where the renormalization scale is taken to be close to the
electroweak {\it vev}.  The trilinear soft terms $A_s,A_s^i$ have been
neglected here since they can be suppressed for reasons mentioned
below (\ref{softops}).  In the above approximation, the CP-even and
the CP-odd neutral Higgs fields are linear combinations of the real
and imaginary parts of $\{H_u^0,\tilde{\nu}_L,S\}$ respectively.  The
charged Higgs $H^+$, on the other hand, is a combination of
$\{H_u^+,\tilde{e}_L^{\dag}\}$.

\subsection{The $\sim$125 GeV Eigenstate}
\label{125}

It is important to understand what region of parameter space of the
model gives rise to an eigenstate with mass near 125 GeV, given the
recent discovery of a Higgs-like particle with that mass.  We will
only make some general and somewhat qualitative comments, leaving a
detailed study of these issues for future work \cite{Kumar:2012yy}.

To start, let us write down the tree level ($\delta \lambda_u=\delta
\lambda_{\nu}=\delta \lambda_3=0$) mass matrix for the CP-even neutral
states in the $(H_u^0,\, \tilde{\nu}_{(a)},\, S_R)$ basis:
\ba
{\cal M}_{H}^2 = \left(\begin{matrix} \frac{1}{2}[(g^2+g'^2)\,v^2 s^2_{\beta} -2 t^{-1}_{\beta} B\mu_L^{(a)}] & 
[-\frac{(g^2+g'^2)}{4}\,v^2 s_{2\beta}+B\mu_L^{(a)}]& v s_{\beta}[\sqrt{2}g' M^D_{1}+2\lambda^S_u(\mu_u+\lambda^S_u\,v_s)] \cr
[-\frac{(g^2+g'^2)}{4}\,v^2 s_{2\beta} +B\mu_L^{(a)}] & \frac{1}{2}[(g^2+g'^2)\,v^2 c^2_{\beta} -2 t_{\beta} B\mu_L^{(a)}] & -\sqrt{2}\,g' v \, c_{\beta} M^D_1 \cr
v s_{\beta} [\sqrt{2}g' M^D_{1}+2\lambda^S_u(\lambda^S_u v_s+\mu_u)] &-\sqrt{2}\,g' v\,c_{\beta} M^D_1 & 
\frac{- 2 (t_s+ \lambda^S_u \mu_u v^2 s^2_{\beta}) + \sqrt{2} g'\,v^2 c_{2\beta} M^D_1}{2 \,v_s}
 \end{matrix} \right)\nonumber
\ea 
where we have used the minimization conditions in (\ref{ewsb-eq}) to
get rid of the non-holomorphic soft mass-squareds for $H_u^0$,
$\tilde{\nu}_{(a)}$ and $S$.  

\begin{wraptable}[17]{r}{0.63\textwidth}
\vspace{-25pt}
\begin{center}
\begin{tabular}{|rcl|rcl|}
\hline
\multicolumn{3}{|c|}{\rule{0mm}{5mm} \bf Benchmark I}			& \multicolumn{3}{|c|}{{\bf Benchmark II}}  \\ [0.3em]
\hline
\multicolumn{3}{|c|}{$\tan \beta = 3$}							& \multicolumn{3}{|c|}{$\tan \beta = 17$}  \\ [0.3em]
\hline
\rule{0mm}{5mm} 
$\lambda^S_u$ & $=$ & $0.1$								& $\lambda^S_u$ & $=$ & $0.1$\\
$\lambda^T_u$ & $=$ & $1.0$ 								& $\lambda^T_u$ & $=$ & $0.9$\\
$\mu_u$ & $=$ & $200$ GeV 								& $\mu_u$ & $=$ & $200$ GeV \\
$M^D_1$ & $=$ & $200$ GeV 								& $M^D_1$ & $=$ & $200$ GeV\\
$M^D_2$ & $=$ & $1000$ GeV 							& $M^D_2$ & $=$ & $1000$ GeV\\
$B\mu_L$ & $\simeq$ & $-(174\,{\rm GeV})^2$ 				& $B\mu_L$ & $\simeq$ & $-(123\,{\rm GeV})^2$\\
$t_S$ & $\simeq$ & $(174\,{\rm GeV})^3$ 					& $t_S$ & $\simeq$ & $(138\,{\rm GeV})^3$\\
$m_{S_R}^2$ & $\simeq$ & $({\rm 1115\,GeV})^2$ 				& $ m_{S_R}^2$ & $\simeq$ & $({\rm 880\,GeV})^2$\\
$m_{T}^2$ & $\simeq$ & $(1450\,{\rm GeV})^2$ 				& $ m_{T}^2$ & $\simeq$ & $(1390 \,{\rm GeV})^2$\\
$m_{\tilde{t}_1}^2 = m_{\tilde{t}_2}^2$ & $=$ & $(500\,{\rm GeV})^2 $ & $m_{\tilde{t}_1}^2 = m_{\tilde{t}_2}^2$ & $=$ & $(500\,{\rm GeV})^2$ \\ [0.3em]
\hline
\multicolumn{3}{|c|}{\rule{0mm}{5mm} $m_h ~\simeq~ 125$ GeV} 								& \multicolumn{3}{|c|}{$m_h ~\simeq~ 125$ GeV}\\
[0.3em]
\hline
\end{tabular}
\end{center}
\vspace{-15pt}
\caption{\small{Two benchmarks giving rise to a lightest CP-even Higgs
mass close to 125 GeV. We take $v_s=-5$ GeV in both cases.}}
\label{higgs-bench}
\vspace{-15pt}
\end{wraptable}
By inspection, one can see that the mixing angle between $S_R$ and
$\{H_u^0,\tilde{\nu}_{(a)}\}$ is essentially controlled by the ratio
$v_s/v$.  Hence, a larger $v_s$ will make this mixing angle larger,
pushing down the lightest eigenvalue due to ``eigenvalue-repulsion".
Thus, $v_s$ should be small in order to maximize the lightest
eigenvalue (we do not consider the possibility of a very light
singlet scalar).  In this limit where the off-diagonal entries are
relatively small, it is then not hard to see that the largest
eigenvalue is predominantly $S_R$, while the $H_u^0-\tilde{\nu}_{(a)}$
block gives rise to a tree-level smallest eigenvalue approaching that
in the MSSM.

Thus, in order to obtain the lightest CP-even Higgs mass around 125
GeV, a reasonably large radiative contribution to the Higgs quartics
(primarily $\delta \lambda_u$) is required.\footnote{As explained
earlier, in this model there are no additional \emph{tree level}
contributions proportional to $(\lambda^S_u)^2$ or $(\lambda^T_u)^2$,
unlike that in \cite{Belanger:2009wf, Benakli:2011kz}.} However,
unlike the MSSM, where the dominant contribution to $\delta \lambda_u$
is provided by the stop squarks, here the adjoints $S$ and $T$ can
also provide a significant contribution through terms proportional to
$\lambda^S_u$ and $\lambda^T_u$ in (\ref{W-final}).  In fact, it is
possible that the bulk of the radiative contribution is provided by
the triplets $T$ in the loop, with $\lambda^S_u$ small and
$\lambda^T_u$ close to unity [see the approximate expressions for
$\{\delta\lambda_u,\,\delta\lambda_{\nu},\,\delta\lambda_3\}$ in
(\ref{deltalambda})].  We give two benchmark examples\footnote{We have
used the full Coleman-Weinberg one-loop effective potential to compute
the lightest CP-even Higgs mass eigenvalue.} in Table
\ref{higgs-bench} with values of the important parameters and for two
choices of $\tan \beta$, close to the minimum and maximum values
in~(\ref{tanbrange}).  In these benchmark examples, the stop squarks
are taken to be around 500 GeV, so the dominant radiative correction
is provided by the triplet $T$.  It turns out that even though the
singlet and triplet scalar masses are $\gtrsim$ TeV, the sensitivity
of the Higgs potential on them is \emph{less} than that on heavy stops
in the MSSM which could generate a $\sim$125 GeV Higgs mass, making
this model significantly less fine-tuned than the MSSM. Part of the
reason is to be found in the factor of 5 versus 3 displayed
in~(\ref{deltalambda}), which makes the contribution to the Higgs
quartic more effective for the triplets than for the stops.  For
instance, for $y_t = \lambda^T_u$, a $500~{\rm GeV}$ ($1~{\rm TeV}$)
triplet gives the same contribution to $\delta \lambda_u$ as stops
with $m_{\tilde{t}_1} = m_{\tilde{t}_2} = 1~{\rm TeV}$
($m_{\tilde{t}_1} = m_{\tilde{t}_2} = 3.2~{\rm TeV}$).  In addition,
one finds that the radiative correction of the triplet to $m^2_{H_u}$
corresponds to that of a \textit{single} stop with $m^2_{\tilde{t}_i}
= m^2_{T}$ and $y_t = \lambda^T_u$.  Hence, for fixed $\delta
\lambda_u$ and equal stop masses (no stop LR mixing), one can estimate
an overall suppression of the triplet ``quadratic divergence" compared
to that of the stops by a factor of about
\ba
\frac{\left. \Delta m^2_{H_u} \right|_{\rm triplet}}{\left. \Delta m^2_{H_u} \right|_{\rm stops}}
&\sim&
\frac{1}{2} \, \frac{(\lambda^T_u)^2}{y^2_t} \, \frac{e^{\frac{16\pi^2 \delta \! \lambda_u}{5(\lambda^T_u)^4}}}{e^{\frac{16\pi^2 \delta \! \lambda_u}{3y_t^4}}}~.
\ea
For $\lambda^T_u \approx 1$ and $m_T \approx 1~{\rm TeV}$
(corresponding to $\delta \lambda_u \approx 0.11$), the above
represents a suppression by a factor of about 20, which results in a
significant reduction in fine-tuning.  We have checked that the
fine-tuning can indeed be mild by computing the logarithmic
derivatives of the EW scale w.r.t.~the microscopic parameters, in the
framework of the 1-loop effective potential.\footnote{We will present
our results in more detail in \cite{Kumar:2012yy}.} Remember also
that, as mentioned earlier, the Higgs potential is much less sensitive
to the (Dirac) gluino mass in this model, compared to the (Majorana)
gluino mass in the MSSM.

It is also interesting to note that a
combined global fit to Higgs properties, ${\rm Br}(B\rightarrow
X_s\,\gamma)$, and the $W$-mass, show a preference for 400-500 GeV
degenerate stops, provided there is an additional mechanism to obtain
a $\sim 125$ GeV Higgs eigenstate \cite{Espinosa:2012in}.  This is
precisely the situation in the benchmark examples in Table
\ref{higgs-bench}, where the triplet provides the additional
contribution to the Higgs mass (its heaviness being motivated by
EWPT).  Finally, note that since the couplings
$\{\lambda^S_u,\lambda^T_u\}$ grow with energy and since $\lambda^T_u$
is close to unity in Table \ref{higgs-bench}, one finds\footnote{We
have computed the RGE's by implementing the full model in
SARAH~\cite{Staub:2008uz,Staub:2010jh}.} a Landau pole at around
$10^7$ GeV ($10^8$ GeV) for Benchmark I (Benchmark II), implying that
additional new physics has to come in around those scales.  This is
consistent with our approach specified in the Introduction; we are
primarily interested in understanding the nature of physics affecting
the LHC, and are agnostic about effects at higher energy scales.
Presumably a microscopic understanding of supersymmetry breaking
within this setup\footnote{At present, we have only done a spurion
analysis of supersymmetry breaking, in Section~\ref{susy-break}.} will
provide insights into the nature of physics at such scales.

Before moving on to discussing various aspects of collider
phenomenology, it is worth commenting on the properties of the
$\sim$125 GeV eigenstate within the model.  For the two benchmarks, it
can be readily checked that this state is primarily a combination of
$H_u^0$ and $\tilde{\nu}_{(a)}$ with only a negligible $S_R$
component.  Thus, this state has properties very similar to the
lightest CP-even state arising within the MSSM (with
$\tilde{\nu}_{(a)}$ replaced by $H_d^0$ of course).  As mentioned at
the beginning of the section, the couplings, production, and decays of
this and the other scalar electroweak states will be studied in detail
in another work \cite{Kumar:2012yy}.

\section{Phenomenology}
\label{pheno}

In this section, we discuss several phenomenological features of the
framework in which lepton number is related to the $R$-symmetry
($R=R_1$).  Since the phenomenology of this class of models is rather
novel, in this paper we only outline the {\it broad} phenomenological
consequences of the framework for collider and neutrino physics.  A
detailed treatment of these issues is provided in a companion paper
\cite{Frugiuele:2012xx}, which studies the existing collider
constraints on this class of models, as well as the various
interesting signals which could be probed in the near future.

The main qualitative features of the phenomenology of this class of
models which sets it apart from traditional supersymmetric models like
the ($R$-parity conserving) MSSM (or many of its cousins like the
NMSSM or models with extra vector-like matter\footnote{These are some
of the popular models which could also give rise to Higgs mass near
125 GeV without much tuning.}), are the following:
\begin{itemize}

\item The existence of a ``Dirac" structure in the gauge sector of the
model.  Among other things,
this gives rise to a suppression of the production of squark pairs at
the LHC compared to the Majorana case, which helps in relaxing the
bounds on superpartners from current searches \cite{Frugiuele:2012xx}.
Another straightforward consequence of this is a suppression of those
signals which depend on the \emph{Majorana} nature of gluinos,
e.g.~same-sign (SS) dileptons.

\item The $R$-symmetry dictates a specific set of operators in the
superpotential and ${\cal L}_{{\rm soft}}$, distinct from the
``standard" cases.  In particular, in the $R = R_{1}$ realization,
there exist ``RPV" operators of the type $\lambda_{ijk} L_i L_j E^c_k$
and $\lambda'_{ijk} L_i Q_j D^c_k$ in the superpotential, and
$B\mu_L^{(i)}\,H_u L_i$ in the soft Lagrangian.  Since these operators
are consistent with the $R$-symmetry (hence with lepton number), they
cannot generate neutrino masses.  Therefore, the sneutrino can have a
significant {\it vev} in these models, thereby acting as a genuine
Higgs field, in stark contrast to standard RPV models.

Also, since the usual trilinear terms involving squark, slepton and
Higgs fields are forbidden, there is no left-right mixing in the
squark and slepton scalar mass-squared matrices.

\item The existence of a sizeable (electron) sneutrino {\it vev}
implies that some of the $\lambda$ and $\lambda'$ couplings are the
lepton and down-type Yukawa couplings, respectively (which are well
known up to $\tan \beta$).  As explained in Sections~\ref{charac} and
\ref{sec:ewp}, this implies that the flavor structure of the $\lambda$
and $\lambda'$ couplings, as well as the various indirect constraints
on these, are rather specific compared to standard RPV models.

Furthermore, it implies that there is mixing between the neutrino(s)
and neutralinos, and between the electron and the charginos, but with
a different dependence on the parameters compared to that in standard
RPV models, as explained in Section~\ref{charac}.

A rich and interesting pattern of signatures results from such a
structure.  For example, decays of the ``LSP"\footnote{``LSP" here
stands for the {\it lightest non SM-like superpartner which is charged
under the SM}.  The ``LSP" is really unstable, just as in RPV models.
Note that the qualification of being charged under the SM is relevant
because the gravitino can be the lightest BSM particle in many cases;
however, in our framework, final states that include a gravitino have
a negligible branching fraction and play no role in collider physics,
{\it unlike} that in gauge mediation (see Section~\ref{distinguish}).
Hence, we will reserve the term ``LSP" for the lightest non SM-like
superpartner charged under the SM, such as a neutralino or stau.  This
is phenomenologically useful since the ``LSP" is the last step of the
SUSY decay chains before producing a pure SM final state.} like
$\tilde{X}_1^{0+} \rightarrow Z \bar{\nu}_e,\, h \bar{\nu}_e,\, W^-
e^+_L$, and $\tilde{\tau}_L^- \rightarrow \tau_R^- \bar{\nu}_e,
\bar{t}_L b_R$ are prompt and have a significant branching ratio,
unlike standard RPV models.

\item The existence of an $R$-symmetry\footnote{It is assumed to be
(explicitly) broken only by a very small amount, so for collider
purposes the symmetry is exact.} implies the conservation of two
charges, the electric charge and the $R$-charge, even after
electroweak symmetry breaking.  In particular, the neutralinos and
charginos are Dirac in nature, and their interactions must conserve
both charges (see Sections~\ref{sec:charginos} and \ref{sec:neutral}).
This gives rise to a rich and interesting pattern of decays of the
neutralinos and charginos, and sleptons and squarks, which is
different from that in the MSSM.

\item In principle, flavor physics can be quite rich as well since
bounds from flavor-violating processes are quite relaxed with an
$R$-symmetry \cite{Kribs:2007ac}.  However, we will not consider this
in detail in this work.
\end{itemize} 
Although different subsets of the above set of signals can be mimicked
by other models, the entire set of signals is rather unique.  Hence,
if the model is correct, it should be possible to distinguish this
class of models from other models in the near future (more about this
in Sections~\ref{LQ-neutrinos} and \ref{distinguish}).

Before going into more details about the phenomenology, it is useful
to have an understanding of the spectrum of the model.  Since the
motivation is to build an electroweak-natural model, we consider a
situation in which the third generation squarks are light ($\lesssim
500$ GeV).  In fact, as will be shown in \cite{Frugiuele:2012xx}, the
bounds on third generation squarks are weaker than 500 GeV, while that
on the first and second generation squarks also turn out to be mild --
in the range 600-700 GeV. This bound assumes a heavy Dirac gluino
($M^D_3\simeq 2$ TeV), which is still consistent with naturalness due
to the existence of a ``supersoft" structure in the gauge interactions
of the model\cite{Fox:2002bu}.  The Dirac wino is expected to be heavy
($\gtrsim$ TeV) to satisfy electroweak precision measurements of the
coupling of the $Z$ to charged leptons for a reasonably large range of
the sneutrino {\it vev} (see Section~\ref{sec:ewp}).  For concreteness
we take $M^D_2\simeq 1.5$ TeV. There are no direct bounds on the Dirac
bino; however anticipating that the origin of its mass is tied to
those of the wino and gluino, we take $M^D_1\simeq 1$ TeV for
concreteness.\footnote{In Section~\ref{125}, we have taken $M^D_1=
200$ GeV, but a CP-even Higgs near 125 GeV is also possible with
$M^D_1=1$~TeV.} Since the $\mu$ parameter is directly connected to
naturalness, we take it around the EW scale, $\mu \simeq 200$~GeV.
This implies that the lightest non SM-like\footnote{Note that since
the neutrino(s) and the electron mix with the neutralinos and charginos
respectively, technically \emph{these} are the lightest neutralino(s) and
chargino, respectively.} charginos and neutralinos are mostly
Higgsino-like.  Finally, the sleptons are expected to be among the
lightest particles in the BSM spectrum because of the close connection
of the slepton sector with EWSB in our framework.  Furthermore, a good
degree of degeneracy among the three generations of sleptons is
expected.  Since the electron sneutrino plays the role of the
down-type Higgs, electroweak naturalness requires its soft mass to be
close to the electroweak scale.  For concreteness, we take
$m_{\tilde{L}}^2 \simeq m_{\tilde{\bar{E}}}^2 \simeq (200-300\,{\rm
GeV})^2$ for all three generations.  Thus, the lightest BSM particles
consist of the sleptons, sneutrinos and Higgsino-like neutralinos and
charginos.  Depending on the situation, either the sleptons/sneutrinos
or the Higgsino-like neutralino could be the ``LSP".

\subsection{Summary of Bounds from Existing Searches}

In Ref.~\cite{Frugiuele:2012xx} we perform a more detailed analysis of
the implications for the LHC of the leptonic $U(1)_R$ symmetry.  We
highlight here a subset of those findings.  In the LHC context, the
small $R$-violating effects can be neglected, and the physics
effectively exhibits a new conserved quantum number, the $R$-charge.
As described above, we focus on the region of parameter space where
the Dirac gaugino masses are in the $1-2$~TeV range, which effectively
decouples (in a first approximation) the bino-singlino,
wino-tripletino and gluino-octetino Dirac states (though the latter
can have some impact on squark pair-production through certain
$t$-channel diagrams).  As a result the lightest (higgsino-like)
neutralino and chargino states play an important role, through the
decays listed in the third item above.  Such decay channels are also
intimately related to the fact that down-type fermions get their
masses from a sneutrino \textit{vev}.  We note that, due to its
higgsino-like nature and the typical branching fractions, the LEP
bounds on charginos are still stronger than those available at the
LHC~\cite{CMS-PAS-SUS-11-016}, in our framework.

Let us start by summarizing the results of the interpretation of the
current ATLAS and CMS searches on the first and second generation
squarks within the leptonic $U(1)_R$ model.  The first point to notice
is that, as emphasized in~\cite{Kribs:2012gx,Heikinheimo:2011fk},
the Dirac nature of the gluino results in a suppression of the squark
pair-production cross section.\footnote{The Dirac gluino
pair-production cross section is larger than in the MSSM, but for
decoupled squarks the current bounds on the gluino mass are only
slightly stronger than for the Majorana case due to the steeply
falling cross section with the Dirac gluino mass.  In any case, we are
taking gluinos at around 2~TeV to emphasize that a spectrum with
squarks lighter than gluinos is fairly natural in our framework.} By
allowing for lighter squarks than in the MSSM, there is an important
secondary consequence that compounds this effect.  Namely, that the
efficiencies of the current searches, which are optimized for
MSSM-like production cross sections, can be significantly reduced due
to the aggressiveness of the present cuts.  The upshot is that squarks
as light as $600-700$~GeV can be consistent with the various generic
SUSY searches (involving jets, varying number of leptons and
$\cancel{E}_T$), to be compared to the current limit of $1.4$~TeV in
the MSSM (when the gluinos are heavy)~\cite{Aad:2012hm}.

We also find that within our framework the bulk of the processes end up
producing some amount of missing energy (in the form of neutrinos), so
that most of the present search strategies can apply with minor
modifications.  However, there are some topologies involving only
visible particles (for instance when the neutralinos decay through
their $We$ channel, with a hadronically decaying $W$).  It would be
interesting to design search strategies for such ``no $\cancel{E}_T$"
channels.  Another important example of no missing energy channel is
the lepto-quark one.  Interestingly, the RH strange squark has a
sizable branching fraction into $\tilde{s}_R \to e^-_L j$.  Current LQ
searches have only a slightly smaller reach than the generic SUSY
searches above (the latter assume degenerate squark masses), and a
discovery in such a channel represents an exciting possibility.
Besides allowing for a measurement of the LQ mass, such decay modes
may allow to extract additional information from ``mixed topologies",
where one of the pair produced particles decays visibly through their
lepto-quark mode, while the other one decays into states involving
missing energy, which can be used for triggering and background
reduction.  One example of this sort is described in
Subsection~\ref{largevev}, and illustrates how one may infer that the
sneutrino \textit{vev} is indeed large, a characteristic of the
leptonic $R$-symmetry.  Other examples are presented
in~\cite{Frugiuele:2012xx}.

The third generation is likely of central importance for understanding
the physics of EWSB. As was emphasized in Section~\ref{125}, the stops
can be reasonably expected to lie in the few hundred GeV range, based
on naturalness considerations, since the bulk of the lightest CP-even
Higgs mass arises instead from the radiative corrections of a heavy
triplet (with moderate associated fine-tuning).  As it turns out, the
lepto-quark nature of $\tilde{t}_L$, $\tilde{b}_L$ and $\tilde{b}_R$
offers a powerful handle into this sector.  As further explained in
the companion paper~\cite{Frugiuele:2012xx}, depending on the
sneutrino \textit{vev}, such searches can easily cover the interesting
expected range.  Still, at present, the strongest constraint arises
from the CMS direct sbottom search~\cite{CMS-summaryplot}, which can
be interpreted as leading to a sneutrino-\textit{vev}-dependent bound
on $\tilde{b}_L$ that varies between $\sim 300$ and $500$~GeV, for
representative values of the model parameters.  The same search
results in a bound on the RH sbottom of close to $500$~GeV. The LH
sbottom bound implies indirectly a lower bound on the LH stop a few
tens of GeV larger, since due to the absence of LR mixing, the stop is
always heavier than the sbottom.  We therefore see that a large part
of the interesting mass range has been tested, and that naturalness
would lead us to conclude that a discovery should be possible in the
relatively near future.  We discuss a further interesting feature of a
possible LQ signal in Section~\ref{LQ-neutrinos}.  One should also
mention that most of the current dedicated searches for third
generation squarks do not apply in a straightforward manner, but it
should be possible to adapt them to cover more general possibilities
than found in simple limits of the MSSM phenomenology.

\subsection{Resonant Slepton/Sneutrino Production}
\label{singleprod}

One of the characteristic features of the presence of the
$\lambda'_{ijk} L_i Q_j D^c_k$ operator is the resonant production of
sleptons and sneutrinos, similar to that in RPV models.  In this
subsection, we would like to study the prospects for resonant
slepton/sneutrino production at the LHC, \emph{assuming} that the
bounds on the $\lambda$ and $\lambda'$ couplings in Tables
\ref{lambda-bounds}, \ref{lambdap2-bounds} and \ref{lambdap3-bounds}
are saturated.  As explained in Section~\ref{sec:ewp}, the sole
exception is the $\lambda'_{311}$ coupling, which we assume to be
negligible ($\simeq 0$).  Furthermore, for simplicity we will study
the simple case where the sleptons and sneutrinos are the ``LSP", so
that the only decay modes of the slepton/sneutrino are via the
$\lambda$ and $\lambda'$ couplings (resonant production occurs via
$\lambda'$ couplings).  Finally, we organize our analysis in terms of
the two-body final states coming from slepton/sneutrino decays.  For a
given final state, we assume that only those $\lambda$ and $\lambda'$
couplings which could give rise to that particular final state are
non-zero and that they saturate the bounds in Section~\ref{sec:ewp}.
For example, for the $e^{\pm}\mu^{\mp}$ final state, this would imply
that the production of both $\tilde{\nu}_{\mu}$ and
$\tilde{\nu}_{\tau}$ is considered turning on those couplings which
allow them to be produced ($\lambda'_{211},\lambda'_{222},
\lambda'_{311},\lambda'_{322}$), and those which allow them to decay
to $e^{\pm}\mu^{\mp}$ ($\lambda_{212},\lambda_{231}$).  Of course, the
$\lambda$ and $\lambda'$ couplings which are identified with the
lepton Yukawa ($\lambda_{122},\lambda_{133}$) and down-type Yukawa
($\lambda'_{111},\lambda'_{122},\lambda'_{133}$) couplings are always
assumed to be present and non-vanishing.\footnote{Note that we assume
that if more than one sneutrino or slepton species can lead to the
given final state, they have comparable masses but are split by an
amount larger than their width, so that interference effects are
negligible.  This is well justified since the widths of the
sneutrinos/sleptons are expected to be extremely small.}
\begin{figure}[t]
\centering
\begin{tabular}{cc}
\includegraphics[width=0.48\textwidth]{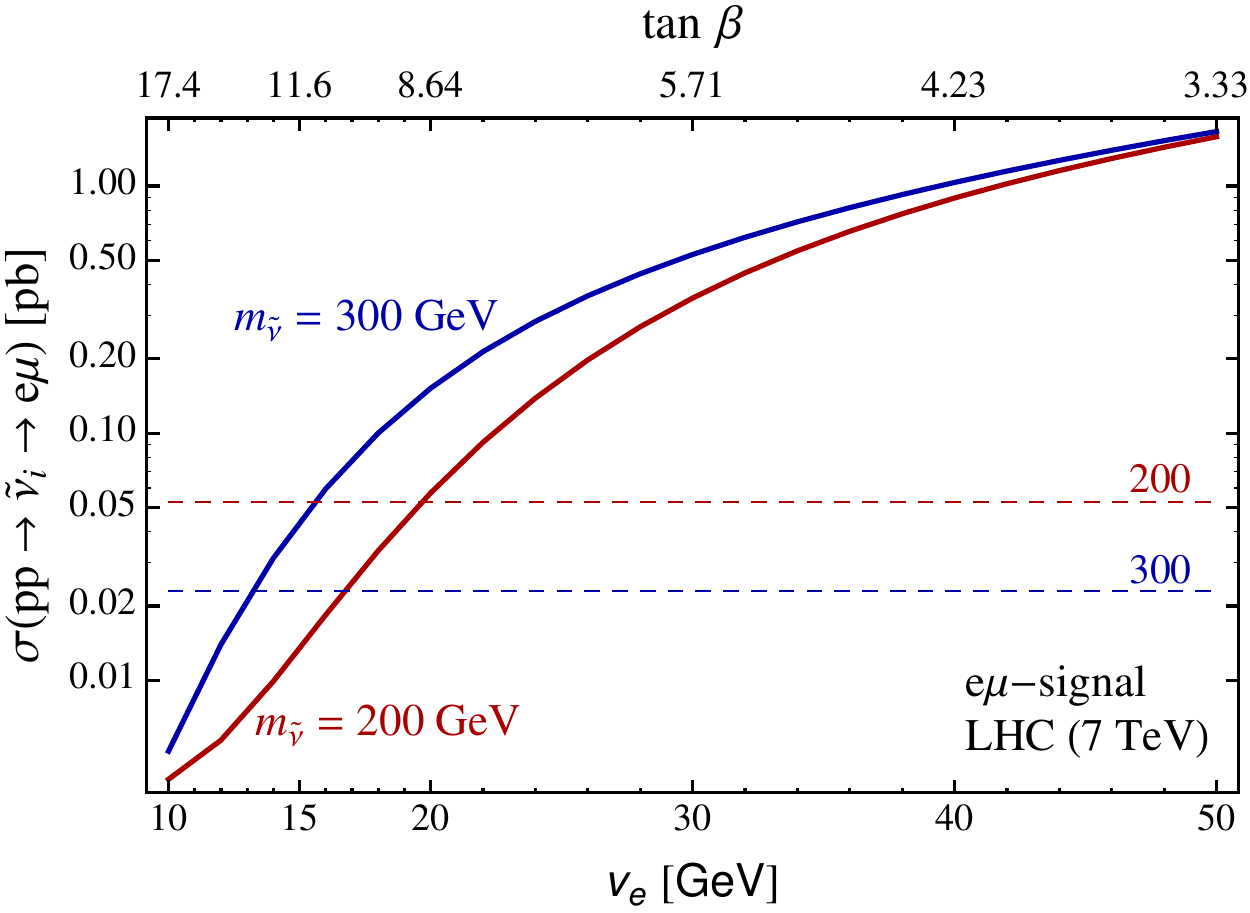} &
\includegraphics[width=0.48\textwidth]{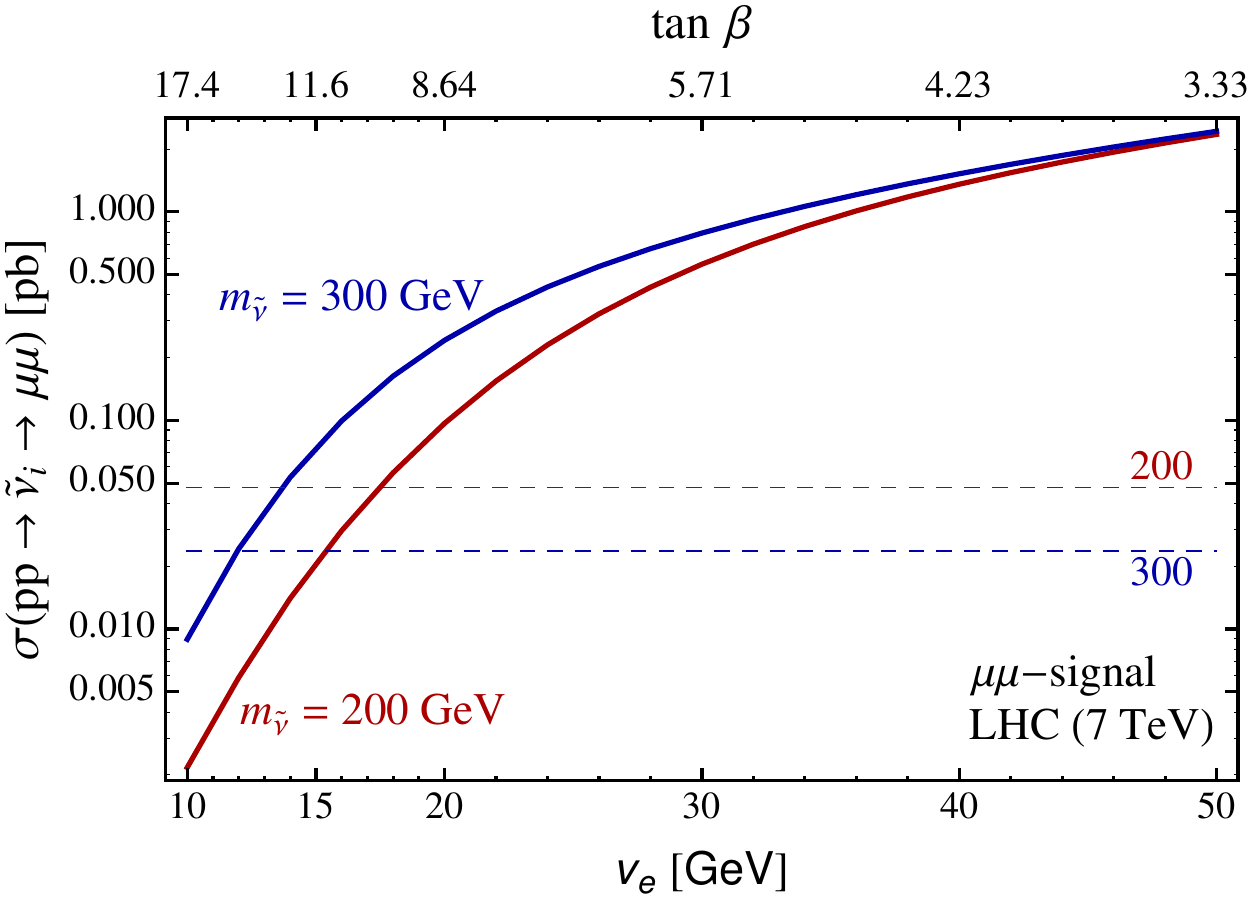}
\end{tabular}
\caption{\footnotesize{Upper bounds on $\sigma_{prod}\times {\rm BR}$
for the $e^{\pm}\mu^{\mp}$ (left panel) and $\mu^+\mu^-$ (right panel)
final states as a function of the electron sneutrino {\it vev} for two
different values of the sneutrino masses: 200 GeV and 300 GeV. For
$e^{\pm}\mu^{\mp}$, both $\tilde{\nu}_{\mu}$ and $\tilde{\nu}_{\tau}$
are produced, and the non-vanishing couplings are taken to be
($\lambda'_{211},\lambda'_{222}, \lambda'_{311},\lambda'_{322}$) for
production and ($\lambda_{212},\lambda_{231}$) for decay.  For
$\mu^+\mu^-$, only $\tilde{\nu}_{\tau}$ is produced, and the
non-vanishing couplings are taken to be
($\lambda'_{311},\lambda'_{322}$) for production and ($\lambda_{322}$)
for decay.  The $\lambda$ and $\lambda'$ couplings identified with
lepton and down-type Yukawa couplings are always present.  The LHC
bounds are shown as dashed lines.  These bounds are provided by
\protect\cite{Jackson:2011zi} for $e^{\pm}\mu^{\mp}$ and by
\protect\cite{Collaboration:2011dca} for
$\mu^+\mu^-$.}}\label{fig:emumumu}
\end{figure}

Comparing with the existing experimental bounds from the various
two-body final states from the Tevatron and the LHC, we find that only
the $e^{\pm}\mu^{\mp}$ and $\mu^+\mu^-$ final states provide
constraints on the parameter space.  Fig.~\ref{fig:emumumu} shows the
bounds on the $\sigma_{prod}\times {\rm BR}$ as a function of the
(electron) sneutrino {\it vev} (varied within the allowed range) for
the $e^{\pm}\mu^{\mp}$ and $\mu^+\mu^-$ final states.  One finds that
only values of the sneutrino {\it vev} close to the minimum value
(maximum for $\tan\beta$) are allowed by the current LHC constraints.
However, since these bounds are for values of couplings
\emph{saturating} the bounds in Section~\ref{sec:ewp}, this simply
suggests that there are good detection prospects for these final
states in the future, within this framework.

\subsection{Lepto-quark (LQ) Signals -- $R$-symmetry at the TeV Scale}
\label{LQ-neutrinos}

Another important consequence of the $\lambda'_{ijk} L_i Q_j D^c_k$
operator is the presence of lepto-quark (LQ) signals.  Again, since
the $LQD^c$ operator is also present in standard RPV models, these
signals are present in principle in RPV models as well.  However, we
will argue below that observation of certain LQ signals will in fact
suggest the existence of an $R$-symmetry in the TeV scale Lagrangian,
which is, furthermore, tied to lepton number ($R=R_1$).  This is made
possible by a connection to neutrino physics, as will be explained.
Finally, we will mention a number of situations in which further
information from other channels would be required in order to rule out
alternative interpretations.  Fortunately, such information should be
accessible at the LHC if the leptonic $U(1)_R$ symmetry is indeed at
play.

Because squarks have $R$-charge 1 in the model, they also carry
lepton-number since it is identified with the $R$-symmetry ($R=R_1$).
These scalar ``lepto-quarks" are pair produced by QCD interactions,
but can decay via the $\lambda'$ coupling above, thereby displaying
their LQ nature.  Such channels can be
very important for the third generation\footnote{The RH strange squark
can also display interesting LQ decay channels,
see~\cite{Frugiuele:2012xx}.} for two reasons -- i) the bounds on
$\lambda'_{333}$ are quite weak from Section~\ref{sec:ewp}, and ii)
the third generation Yukawa couplings\footnote{Recall that when
$R=R_1$, the couplings $\lambda_{1jj}$, $j=2,3$, and $\lambda'_{1kk}$,
$k=1,2,3 $, are identified with the lepton and down-type Yukawa
couplings, respectively.} are the largest, while the third generation
squarks can naturally be the lightest.  In particular, this means that
$\lambda'_{133}$ and $\lambda'_{333}$ can be ${\cal O}(1)$, and could
give rise to the decays
\ba
\label{LQ-decay}
\tilde{t}_L &\rightarrow& b_R\, l^+_L~,
\hspace{5mm}
\tilde{b}_R ~\rightarrow~ t_L\, l^-_L~,
\hspace{5mm}
l_L ~=~ e_L,\,\tau_L~,
\nonumber \\ [0.4em]
\tilde{b}_L &\rightarrow& b_R \,\bar{\nu}~,
\hspace{6.5mm}
\tilde{b}_R ~\rightarrow~ b_L\, \nu~,
\hspace{7.2mm}
\nu ~=~ \nu_e,\,\nu_{\tau}~,
\ea
plus their conjugate processes.  Since for an electroweak-natural
model we expect the third generation squarks to have masses $\lesssim$
500 GeV, there are good prospects of observing these signals.  The
observation of the LQ signal (\ref{LQ-decay}) may provide support for
the existence of the {\it $U(1)_R$ symmetry in the full Lagrangian at
the TeV scale}, which is a stronger conclusion than the one one could
reach based on the observation of Dirac gluino signatures alone.  The
argument is the following.

Let us start by assuming that some of the LQ signals (\ref{LQ-decay})
have been observed [either from $\tilde{b}_R$ or from the
$(\tilde{t}_L,\tilde{b}_L)$ doublet].  Although it may be hard to
distinguish the decays in the second line in (\ref{LQ-decay}) from the
``standard" SUSY decays involving a (massive) neutralino instead of
the neutrino, the simultaneous observation of the (fully visible)
decays involving a charged lepton may be taken, based on $SU(2)_L$
invariance, as indication that (at least part of) the missing energy
signal is associated with a neutrino.  We will also assume that the
observed LQ is indeed a \textit{third generation} squark (as opposed
to first/second generation), and will further comment on this
assumption below.

\begin{wrapfigure}[8]{r}{0.4\textwidth}
\vspace{-25pt}
\centering
\includegraphics[width=0.35\textwidth]{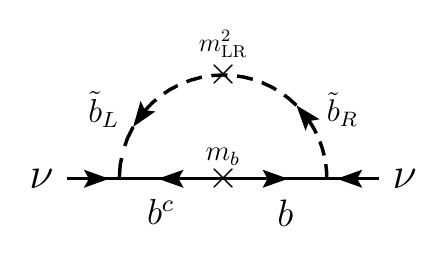}
\vspace{-25pt}
\caption{\footnotesize{Contribution to the neutrino mass from a SUSY
lepto-quark.}}
\label{fig:Neutrino-LR}
\vspace{-35pt}
\end{wrapfigure}
Given the above LQ observation, it is possible to obtain a
surprisingly large amount of information regarding the structure of
the soft supersymmetry breaking terms.  This is because the $LQD^c$
operators lead to a contribution to the neutrino masses given by:
\ba
\Delta m_{\nu_i} &\sim& \frac{3(\lambda'_{i33})^{2}}{16\pi^{2}} \frac{m_{\rm LR}^2}{m^{2}_{\tilde{b}}}~m_{b}~,
\hspace{1cm}
i=1,2,3~,
\label{AbBound}
\ea
where $m_{{\rm LR}}^2$ is the left-right mixing in the sbottom
mass-squared matrix.  Note that in the $R$-symmetric limit 
this left-right mixing is forbidden, but it is present in
RPV models in general.  For an electroweak-natural model, barring
fine-tuned cancellations amongst several contributions, there are two
ways in which Eq.~(\ref{AbBound}) can be consistent with the upper
bound on neutrino masses: $i)$ a suppressed coupling $\lambda'_{i33}$,
or $ii)$ a suppressed $m^2_{\rm LR}$.

In RPV models, where typically $m^2_{\rm LR} = m_b (A_b - \hat{\mu}
\tan\beta) \sim {\cal O}(m_b m_{\tilde{b}})$, this would imply
situation (i) - a very suppressed coupling $\lambda'_{i33}$ ($\lesssim
10^{-3}$).  Here $\hat{\mu}$ denotes the traditional
``$\mu$-term".\footnote{More precisely, the relevant $\hat{\mu}$-term
corresponds to the superpotential bilinear linking the Higgs
doublets giving rise to up-type and down-type fermion masses, and is
\emph{not} the same as $\mu$ in (\ref{W-final}).  In fact, in
$R$-symmetric models, $\hat{\mu}=0$.} However, the observation of a LQ
signal with a very suppressed $\lambda'_{i33}$ coupling is only
possible for a \emph{rather special} situation, i.e.~when the LQ decay
channel (controlled by $\lambda'_{i33}$) of the relevant squark
(sbottom here) has a significant branching ratio (BR) in spite of
$\lambda'_{i33}$ being very suppressed.  We will discuss cases in
Section~\ref{special} when this special situation is satisfied.  In a
typical situation, however, one expects at least some neutralinos
and/or charginos to be lighter than the third generation squarks
(which could have masses $\sim 400-500~{\rm GeV}$).  This is also
motivated by electroweak-naturalness, where a natural solution of the
EWSB minimization conditions typically requires $\mu \lesssim 200-300$
GeV,\footnote{$\mu$ is the coefficient of the term $H_u R_d$ in the
superpotential in (\ref{W-final}).} implying the existence of at least
one neutralino and one chargino lighter than the squark (sbottom
here).  Then standard SUSY two-body decays of the squarks are also
open, and can be used to set a lower bound on $\lambda'$, based on the
fact that the BR associated with the lepto-quark channel cannot be
very suppressed if such a signal is seen at the LHC.

In Appendix \ref{lambdap-bound}, we estimate this lower bound on
$\lambda'_{i33}$ for the two final states in (\ref{LQ-decay}) without
missing energy - (i) charge $\frac{2}{3}$, $l_L^+\,b_R$, and (ii) charge
-$\frac{1}{3}$, $l_L^-\,t_R$.  For (i), we find $\lambda'_{i33}\gtrsim
0.01-0.1$, while for (ii) the lower bounds are slightly smaller.
Thus, for the charge $\frac{2}{3}$, $l_L^+\,b_R$ final state, we
estimate:
\ba 
\label{LR-upper}
m^2_{\rm LR} &\lesssim& (0.005-0.05) \left(\frac{m_{\tilde{b}}}{500~{\rm GeV}} \right)^2 ~{\rm GeV}^2~,\nonumber\\
A_b - \hat{\mu} \tan\beta &\lesssim& (0.002-0.02)  \left(\frac{m_{\tilde{b}}}{500~{\rm GeV}} \right)^2~{\rm GeV}~,
\label{LRBound1}
\ea
while the charge -$\frac{1}{3}$, $l_L^-\,t_R$ final state could be
consistent with a sbottom LR mixing term about two orders of magnitude
larger.  This implies that barring fine-tuned cancellations, both
$A_b$ and $\hat{\mu}$ must be highly suppressed relative to
$m_{\tilde{b}}\sim M_{{\rm SUSY}}$.  Furthermore, note that the upper
bounds (\ref{LR-upper}) are valid at around the electroweak scale.
Now, $A_b$ in particular gets contributions from RG running from
all Majorana gaugino masses as well as other $A$ terms according to:
\ba
\frac{d A_b}{d t} &\simeq& \frac{1}{16\pi^2} \left\{ \frac{32}{3} g^2_3 M_3  + 6 g^2_2 M_2 + \frac{14}{15} g^2_1 M_1+ 2 y_t^2 A_t + 12 y^2_b A_b \right\}~,
\ea
where $y$ and $A$ (with appropriate subscripts) are the Yukawa
couplings and $A$-terms,\footnote{The trilinear soft terms are defined
with one power of the corresponding Yukawa coupling factored out, as
is customary in the flavor diagonal case.} respectively, while $M_i$
are the $SU(3)_C \times SU(2)_L \times U(1)_Y$ Majorana gaugino
masses.  Even for a low scale cutoff $\Lambda \sim 10~{\rm TeV}$, the
individual contributions to $A_b$ would be:
\ba
\Delta A_b &\sim& \{0.2 \, M_3, 0.05 \, M_2, 0.002 \, M_1\}~{\rm GeV}~,
\hspace{3.43cm}
(\textrm{from Majorana masses})~,
\nonumber
\ea 
%
%
\ba
\Delta A_b &\sim& \{0.04 \, A_t, 6\times 10^{-4}\, A_b, 4\times 10^{-5} \, A_\tau\}  \left(\frac{\tan\beta}{3}\right)^2~{\rm GeV} ~,
\hspace{0.2cm}
(\textrm{from A terms})~.
\ea 
The weaker bound in (\ref{LR-upper}) would then imply
\ba
(M_3, M_2, M_1) &\lesssim& (0.1, 0.4, 10) \left(\frac{m_{\tilde{b}}}{500~{\rm GeV}} \right)^2~{\rm GeV}~,\nonumber\\
(A_t, A_b, A_\tau) &\lesssim& (0.5, 30, 500)  \left(\frac{m_{\tilde{b}}}{500~{\rm GeV}} \right)^2 \left(\frac{3}{\tan\beta}\right)^2~{\rm GeV}~,
\vspace{-10pt}
\ea 
which would become stronger by one order of magnitude if the stronger
bound in (\ref{LR-upper}) applies.  Thus, in addition to upper bounds
on $A_b$ and $\hat{\mu}$ from (\ref{LR-upper}), we also get
significant upper bounds on the \textit{three} Majorana gaugino masses
as well as on $A_t$.  For larger $\tan\beta$ the bounds are even
tighter and one could also infer an upper bound on $A_\tau$ as low as
a few GeV. Since the absence of Majorana gaugino masses, $A$-terms,
and the $\hat{\mu}$ term are the hallmark of an $R$-symmetry, the
observation of a lepto-quark signal, via a connection to neutrino
masses, allows one to build a strong case for the presence of an
approximate $U(1)_R$ symmetry at the TeV scale.

\subsubsection{Special Cases}
\label{special}

As mentioned earlier, there exists special cases in which RPV models
can also give rise to visible LQ signals, consistent with neutrino
masses and other constraints.  First, it may be possible for the LQ to
correspond to a second generation squark, e.g.~$\tilde{c}_L$, decaying
via the $\lambda'_{123}$ coupling as $\tilde{c}_L \to b_R e^+_L$ (and
$\tilde{s}_L \to b_R \bar{\nu}_e $).  The point is that such couplings
are not constrained by neutrino mass bounds, and the constraints arise
from very loose ``single coupling bounds", or from ``product coupling
bounds" that can allow $\lambda'_{123}$ to be sizeable if the other
coupling is sufficiently suppressed.  If couplings like
$\lambda'_{123}$ are indeed sizeable, special kinematic configurations
would not be required for the LQ channel to have a visible branching
fraction.  Nevertheless, observation of a LQ signal is only feasible
if the LQ is well below 1~TeV. Thus, given the strong bounds on first
two generation squarks in RPV models without an $R$-symmetry, such an
interpretation would require a significant mass splitting between
$\tilde{c}_L$ and the other squarks.  Such a situation would be
distinguishable from the case where the LQ signal arises from third
generation squarks.  For instance, the ``standard" decays of the LQ to
the ``LSP" (which decays further), for example $\tilde{c}_L \to c
\tilde{\chi}^0_1$ versus $\tilde{t}_L \to t \tilde{\chi}^0_1$, could
distinguish between the two situations.

Thus, we focus on the more natural SUSY interpretation of a LQ signal
as a third generation squark, which might still be consistent with a
very suppressed $\lambda'_{i33}$.  This happens when {\it all other}
decay modes of the squarks are suppressed, so that the LQ decay modes
have a significant branching fraction even with suppressed
$\lambda'_{i33}$ couplings.  Some situations in which this can arise
are the following:
\begin{itemize}
\vspace{-10pt}
\item
There is only a single LQ signal and it is the LSP. Then \emph{only}
LQ decay modes are available, if the LQ is $\tilde{b}_R$.  If the LQ
is the doublet, one could hope to use the decay of the heavier into
the lighter $SU(2)_L$ component, via an (off-shell) $W$, to extract
additional information.  However, besides the 3-body phase space
suppression factor, the corresponding partial width scales like
$(\Delta m)^5$, where $\Delta m$ is the splitting between the two
$SU(2)$ lepto-quark components.  As a result, the derived lower bound
on $(\lambda')^2$ is not useful to conclude that the LR mixing should
be suppressed.

\item The observed LQ is $\tilde{b}_R$ while the doublet LQ's are too
heavy, and/or the BR in their LQ channels is too small (so that they
are not seen in those channels), and if in addition the LSP neutralino
is almost pure wino or almost pure $\tilde{H}_u$, then the 2-body
decay of $\tilde{b}_R$ into the neutralino LSP is highly suppressed.
This case could then be similar to the previous one, unless the
second-lightest neutralino is lighter than $\tilde{b}_R$.

\item
All neutralinos and charginos are heavier than the LQ (or very near
threshold), even if the LQ is not the LSP. One then has to compare
against loop-induced 2-body decays, or 3-body decays.  Such decays are
sufficiently suppressed that a sizeable BR in the LQ channel is
allowed, consistent with neutrino bounds and without LR suppression.

\item Even though the LSP is a neutralino, the LQ happens to be a
(highly mixed) stop with the $t\chi^0$ channel kinematically closed.
One is then again left with loop-induced 2-body decays, or 3-body
decays that can compete against the LQ signal even for rather small
$\lambda'$~\cite{Das:2005mr}.
\vspace{-5pt}
\end{itemize}
In order to further discriminate between the $U(1)_R$-symmetry
interpretation and the above possibilities, further information
regarding the SUSY spectrum would be required.  For example, the
observation of the prompt decay of the neutralino to say $W e$, which
indicates the presence of an appreciable sneutrino {\it vev}, could be
used to reconstruct the mass of the neutralino as well as shed light
into the structure of neutralino interactions, providing support for
this class of models over traditional RPV models.  A thorough
discussion of these issues is left for future work.

\subsection{LHC Signals of a Large Sneutrino {\it vev}}
\label{largevev}

We have seen in previous subsections that the model produces several
distinctive signals at the LHC. Although standard RPV models can give
rise to many of these signals in principle, we saw in
Sections~\ref{singleprod} and \ref{LQ-neutrinos} that the full pattern
of signals is \emph{generically} different.  Furthermore, there is a
particular signal topology which clearly distinguishes this model from
standard RPV models. Not surprisingly, this difference arises due
to the presence of a significant sneutrino {\it vev}, which is a
distinctive feature of the model.

Such a sneutrino {\it vev} opens up the neutralino decay modes into $
W e, Z \nu $ and $ h \nu$.  As will be explained in detail in the
companion paper \cite{Frugiuele:2012xx}, these decay modes have
sizeable branching fractions when $ \tilde X^{0+}_1$ is the ``LSP".
We therefore focus on this particular situation.  The decay mode into
$ W e$ points unambiguously toward a mixing between the electron and
the charginos.  Although this signal could be interpreted also in the
context of a standard RPV model with the left-handed sneutrino
acquiring a {\it vev}, since the bounds from neutrino masses on the
sneutrino {\it vev} are so stringent in the standard case, the
neutralino typically decays outside the detector or through a
displaced vertex.  Therefore, a {\it prompt} decay of the neutralino
into $ W e$ is a clear sign of a sizable sneutrino \textit{vev} and
therefore a hint of a leptonic $R$-symmetry ($R=R_1$).

Furthermore, even in the case where the decay $\tilde{X}_1^{0+}
\rightarrow Z \bar{\nu}_e$ or $h \bar{\nu}_e$ is dominant (with the
$We$ channel suppressed, as can happen in some regions of parameter
space), the observation of a ``mixed topology" signal\footnote{That
is, a LQ decay channel on one cascade leg and a ``standard" SUSY decay
on the other one.} from the pair-production of third generation
squarks, could provide a large amount of information.  For example if
one of the pair-produced $\tilde{t}_L$'s decays via a LQ
channel\footnote{This has a significant branching ratio, especially if
$\lambda'_{333}$ saturates the bounds in Section~\ref{sec:ewp}.} as in
Section~\ref{LQ-neutrinos}, while the other one decays via
$\tilde{t}_L \rightarrow t\,\tilde{X}_1^{0+} \rightarrow
t\,\{Z\bar{\nu}_e,\,h\bar{\nu}_e\}$, this could be argued to provide
evidence for a leptonic $R$-symmetry.  This is because an observation
of such a signal would allow us to draw several conclusions:
\begin{itemize}
\item First, there is a neutralino lighter than $\tilde t_L$.  In
addition, the $\lambda'$ coupling of the LQ decay channel is large
enough to give an observable signal.  In particular, one can conclude
that the magnitude of $ \lambda'$ is comparable to that of the
electroweak gauge couplings.  \item The invisible particle is most
probably a neutrino.  It cannot be a neutralino LSP (in an RPV-MSSM
scenario) since it would have decayed promptly through the $\lambda'$
coupling above into $b$-quarks and a neutrino (but we are imagining
that the $Z$ or $h$ above have been reconstructed).  The invisible
particle cannot be the gravitino either since the three body decay
mode of the neutralino into $ b \bar b \nu $ via a $\lambda' $
comparable to electroweak gauge couplings will dominate over the two
body decay into $ \tilde{G} Z $ or $ \tilde{G} h$.  For example, for a
bino-like NLSP, $ \Gamma (\tilde{\chi}_1^0 \rightarrow \tilde{G}Z) =
\frac{m_{\chi_1}^5}{48\pi m_{3/2}^2 M_{pl}^2} \left(
1-\frac{M_Z^2}{m_{\chi_1}^2} \right)^4 \sin^2{\theta_W}$, while $
\Gamma( \tilde{\chi}_1^0 \rightarrow b \bar{b} \nu ) \approx
\lambda'^2 g'^2 \frac{7m_{\chi_1}^5}{12288\,\pi^3 m_{\tilde b}^4} $
where $m_{\tilde b}$ is the mass of the (off-shell) sbottom squark.
For $\lambda'$ comparable to electroweak gauge couplings and
$m_{\tilde{b}}$ less than a few TeV, the decay width into the
gravitino is always suppressed compared to the three body decay.

\item Finally, as argued earlier, an observation of the
\textit{prompt} decays $\tilde{\chi}_1^{0} \rightarrow Z\nu,\,h \nu$
is strongly disfavored in standard RPV models from bounds on neutrino
masses.

\end{itemize}
Therefore, an observation of the above mixed topology would also
strongly point towards a large sneutrino {\it vev}, and hence towards
a leptonic $R$-symmetry ($R=R_1$).

\vspace{-3mm}
\subsection{Distinguishing from Other Models}
\label{distinguish}

We now briefly discuss how some other popular supersymmetric models
can be distinguished from this model.  The ``classic" and well studied
cases of the ``constrained MSSM (cMSSM)" and ``phenomenological MSSM
(pMSSM)" are easy to distinguish from this model.  This is because the
presence of $R$-symmetry in both the gauge and matter sectors, implies
many signals which are quite different from the cMSSM and pMSSM
models, such as:
\begin{itemize}
\vspace{-5pt}
\item[(a)] Different production rates and decay modes for squarks.

\item[(b)] Dirac gauginos with suppression of SS dileptons signals. 

\item[(c)] Resonant slepton/sneutrino production. 
 
\item[(d)] Lepto-quark (LQ)
signals.  

\item[(e)] Decay of the ``LSP" giving rise to fewer channels with
$\slash{\!\!\!\!E}_T$, etc.
\vspace{-5pt}
\end{itemize}
The issue of distinguishing GMSB models from this model is more
interesting.  In ``standard" GMSB models (without any $R$-symmetry),
signals (a)-(d) above should still allow us to easily distinguish it
from this model.  In $R$-symmetric gauge mediation models (but with
the $R$-symmetry {\it not} related to lepton number, i.e.~$R\neq
R_1$), the signals (a) $\&$ (b) are the same as in this model.
Signals (c) $\&$ (d), however, arise from the $LQD^c$ operators which
are only present when $R=R_1$; hence signals (c) and (d) can be used
to distinguish among the two models.

Signal (e) deserves some more comments.  In GMSB, the gravitino is the
LSP, hence the NLSP (such as a neutralino or a stau) can decay to the
gravitino via $\tilde{\chi}_1^0 \rightarrow Z \tilde{G},\, h\tilde{G}$
and $\tilde{\tau} \rightarrow \tau\tilde{G}$ \cite{Kats:2011qh}.  This
can resemble {\it some} of the decays of the ``LSP" in our model:
$\tilde{X}_1^{0+} \rightarrow Z \bar{\nu}_e,\, h \bar{\nu}_e$ and
$\tilde{\tau}^-_L \rightarrow \tau^-_R \bar{\nu}_e$, as explained at
the beginning of Section~\ref{pheno}.  However, within our framework,
the decay modes $\tilde{X}_1^{0+} \rightarrow W^- e^+$,
$\tilde{\tau}^-_L \rightarrow \bar{t}_L b_R$, etc., can have a
non-trivial branching fraction \cite{Frugiuele:2012xx}, and do not
give any $\slash{\!\!\!\!E}_T$ when the $W$'s decay fully
hadronically, thus allowing for a full reconstruction.  Therefore,
these decays can clearly distinguish our model from GMSB models.  Note
that the gravitino is, of course, also present within our model, and
is also very light.  However, due to its suppressed couplings with
matter, the branching ratios of the ``LSP" to final states with a
gravitino are extremely small and can be neglected.\footnote{Note that
this is in contrast to standard GMSB models, and even to $R$-symmetric
GMSB models with $R=R_0$, where the \emph{only} available decay mode
of the NLSP is to final states with a gravitino.}

\subsection{The Case $``B=R"$}
\label{b-r}

We briefly mention some phenomenological features of the case $R=R_2$,
i.e.~in which the baryon number is identified with an $R$-symmetry.
As mentioned earlier, this case has already been studied in
\cite{Brust:2011tb} and some aspects of the collider phenomenology
have been studied in detail in \cite{Brust:2012uf}.

In these models, the baryonic RPV operators $\lambda''_{ijk} U^c_i
D^c_j D^c_k$ are consistent with the $R$-symmetry.  Therefore, the ``LSP"
decays into jets via the $\lambda''$ couplings in this case, giving
rise to final states with very small missing energy.  Thus, they may
naturally evade most of the current LHC bounds, and have generated
renewed interest.  However, it is important to note that there is a
subtlety.  In order to evade the bounds, the decay of the ``LSP" has to
be prompt, which puts a lower bound on the $\lambda''$ couplings,
$\lambda'' \gtrsim 10^{-5}-10^{-4}$\cite{Barbier:2004ez}.  However,
the indirect bounds on many of these couplings from
neutron-antineutron ($n-\bar{n}$) oscillations are orders of magnitude
stronger than this \cite{Barbier:2004ez}!\footnote{This is true if the
couplings are ``flavor-generic".} Thus, having a large enough coupling
requires either a specific flavor ansatz \cite{D'Ambrosio:2002ex}, or
some other mechanism which relaxes the bounds, for example if
$R$-parity is broken collectively \cite{Ruderman:2012jd}.

\begin{wrapfigure}[16]{r}{0.5\textwidth}
\vspace{-10pt}
\centering
\includegraphics[width=0.37\textwidth]{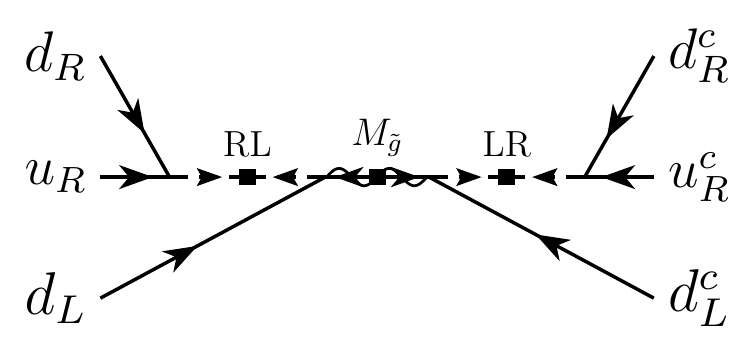}
\vspace{6mm}
\includegraphics[width=0.24\textwidth]{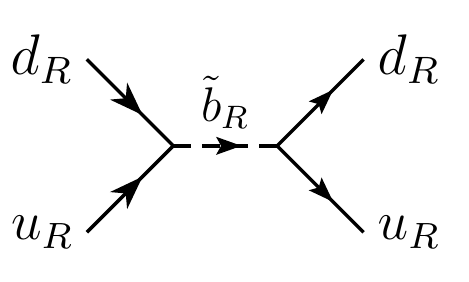}
\vspace{-25pt}
\caption{\footnotesize{Upper panel: $n-\bar{n}$ oscillations from
sbottom exchange via $\lambda''_{113}$, showing the Majorana gluino
mass and LR insertions (as boxes).  The arrows indicate the $R$ number flow.
Lower panel: single sbottom resonant production via~$\lambda''_{113}$.}}
\label{fig:nnbar-diquark}
\vspace{-35pt}
\end{wrapfigure}

Identifying the baryon number with an $R$-symmetry provides an elegant
way to evade these bounds (this point was already made in
\cite{Brust:2011tb}).  Indeed, analogous to the $L=R$ case, here the
$U^c D^c D^c$ operator is consistent with the $R$-symmetry which is
identified with the baryon number; hence there are no bounds on
$\lambda''$ couplings from baryon number violating processes like
$n-\bar{n}$ oscillations.  The baryonic $R$-symmetry will be broken by
the gravitino mass $m_{3/2}$, hence $n-\bar{n}$ oscillations will
constrain $m_{3/2}$ depending on the details of $R$-breaking and
mediation.  For example, the contribution to $n-\bar{n}$ oscillation
from tree level sbottom/gluino exchange (see
Fig.~\ref{fig:nnbar-diquark}) provides an upper bound on
$\lambda''_{113}$~\cite{Barbier:2004ez}:
\ba \lambda''_{113} \lesssim 1 \times \left(\frac{2\,{\rm
GeV}}{m_{LR}}\right)^2\,\left(\frac{2\,{\rm
GeV}}{M_{\tilde{g}}}\right)^{1/2}\,\left(\frac{M^D_3}{1\,{\rm
TeV}}\right)\,\left(\frac{m_{\tilde{b}}}{500\,{\rm GeV}}\right)^4~,
\ea
where $M_{\tilde{g}}$ is the Majorana gluino mass and $m_{LR}$ is the
left-right mixing among the sbottom squarks.  Thus, an ${\cal O}(1)$
$\lambda''$ coupling is possible for $M_{\tilde{g}},m_{LR} \sim$ few
GeV. Within generic gravity mediation, this implies $m_{3/2} \sim$ few
GeV while for anomaly mediation $m_{3/2} \sim 100$ GeV. If $m_{3/2} >
m_{proton}$, then there is also no constraint from proton decay (to
the gravitino, since in that case it would be the lightest baryon!).
In Fig.~\ref{fig:nnbar-diquark} we also show how the $n-\bar{n}$
oscillation process is related to the single resonant production of
$\tilde{b}_R$.  Such a diquark signal would formally play the
analogous role of the lepto-quark signals in the $L=R$ realization
discussed in previous sections.  It would be interesting to try to use
this feature to infer that the TeV scale Lagrangian does indeed
display an approximate $R$-symmetry, as was done in the $L=R$ case.
This is significantly harder here, since the $n-\bar{n}$ constraint is
intimately connected to the first generation, while establishing that
the jets in a diquark signal are in fact connected to the first rather
than the second generation is not quite feasible.  Of course, even
observing such a resonance is significantly more challenging than
observing a LQ signal, especially below the 1~TeV scale (most recent
LHC di-jet searches impose cuts above 1~TeV to control the QCD
backgrounds).  In this sense, a diquark signal arising from the first
two generation squarks (which can be heavier consistent with
electroweak naturalness), through $\lambda''_{111}$ or
$\lambda''_{112}$, can be interesting.  However, since the squarks
could well lie below 1~TeV in this model, the low mass region should
be kept in mind for such searches.  This would also apply for third
generation squarks decaying into dijets.

\section{Dark Matter}
\label{dm}

Since in this class of models the ``LSP" decays, there is no WIMP Dark
Matter (DM) candidate, similar to generic RPV models.  However, other
mechanisms for DM generation are available.  We leave a detailed
exploration of these issues for future work, and only briefly mention
a few possibilities.

First, even in the minimal setup considered in this paper, the
gravitino can provide a natural DM candidate, as follows.  Although we
have not studied effects of $R$-breaking in this paper in detail, such
operators are necessarily present, as explained in Appendix
\ref{r-breakops}.  One of the crucial consequences of these
$R$-breaking operators is the generation of neutrino masses as
mentioned earlier.  Since the scale of $R$-breaking is ultimately tied
to $m_{3/2}$, the upper bound on neutrino masses places an upper bound
on $m_{3/2}$, the precise magnitude of which depends on the details of
$R$-breaking mediation (see Appendix \ref{r-breakops} for a discussion
on two natural possibilities - i) generic gravity mediation, and ii)
anomaly mediation).  Ref.~\cite{Bertuzzo:2012su} has studied fitting
the entire pattern of experimentally measured neutrino masses and
mixing angles within this framework, and finds that $m_{3/2} \lesssim
{\cal O}({\rm keV})$ for generic gravity mediation, and $m_{3/2}
\lesssim {\cal O}(1-100)\,{\rm MeV}$ for anomaly mediation.  Assuming
that the initial gravitino abundance is negligible and is such that it
never reaches thermal equilibrium,\footnote{These conditions can be
easily satisfied since the gravitino has extremely suppressed
couplings.} gravitino masses in the keV-MeV range can indeed provide
the DM abundance of the Universe by one of the following two
processes:
\begin{itemize}
\vspace{-5pt}
\item Thermal scattering of superpartners in the early Universe, as
originally explained in \cite{Moroi:1993mb}.  In this case, the
gravitino abundance depends linearly on the reheat temperature of the
Universe $T_R$.  \item Decays of superpartners, which are still in
thermal equilibrium, to gravitinos -- also known as {\it Freeze-in}
(FI) \cite{Hall:2009bx}.  
In this case, the gravitino abundance is independent of $T_R$.
\vspace{-5pt}
\end{itemize} 
Depending on the superpartner spectrum, one or the other process may
dominate or they may both be comparable.  Gravitino FI has been
studied in \cite{Cheung:2011nn} (for other discussions of the FI
mechanism, see \cite{Cheung:2010gj,Cheung:2010gk,Hall:2010jx}).  An
important point to remember is that since technically the neutrino(s)
are the lightest neutralinos in our model (by virtue of lepton number
being an $R$-symmetry, see Section~\ref{sec:neutral}), the gravitino
is \emph{not} absolutely stable.  The dominant decay mode of the
gravitino $\tilde{G}$ is the process: $\tilde{G} \rightarrow
\gamma\,+\,\nu$, and it can be shown that the gravitino lifetime is
sufficiently long so as to satisfy all observational
constraints\footnote{We disagree with earlier results for the decay
width of the process $\tilde{G} \rightarrow \gamma\,+\,\nu$.  We
believe this discrepancy arises due to the earlier works not taking
into account gauge invariance properly, in particular due to
effectively using a \emph{non-gauge-invariant} regulator.}
\cite{Gregoire:2012zz}.

Finally, it is worth commenting on the tentative $\gamma$-line signal
at around 130 GeV from the Galactic Center (GC) observed by many
groups \cite{Bringmann:2012vr,Weniger:2012tx,Tempel:2012ey,Su:2012ft} in the
FERMI-LAT data.  It is clear that if the signal turns out to be
correct (confirmed by FERMI-LAT), \emph{and} if it is attributed to DM
annihilation, then this cannot be explained within the framework
above, at least in its minimal incarnation.  Explaining the signal
from DM annihilation within this framework would presumably require
the existence of an appropriate additional (dark) sector.  Studying
these issues is left for future work.

\section{Conclusions and Future Directions}
\label{conclude}

Without a doubt, we have entered a data-rich era that is expected to
finally unravel the mystery of electroweak symmetry breaking,
i.e.~uncover the physical microscopic mechanism underlying this
well-established phenomenon.  Although the discovery of a Higgs-like
particle near 125 GeV and the absence of new physics so far provides
some support for electroweak-tuned theories, it is still rather
premature to abandon electroweak-naturalness.  Indeed, Nature could be
cleverly realizing an electroweak-natural model which manifests itself
at the LHC in non-standard ways.  In this work, we have studied one
such elegant model, the defining feature of which is the existence of
a continuous $R$-symmetry, \textit{that coincides with lepton number
when restricted to the SM sector}.  {\it An important consequence of
this is that one of the left-handed sneutrinos gets a significant {\it
vev}, which is not constrained by the upper bound on neutrino masses}
(we take the sneutrino getting a {\it vev} to be of the electron type,
so that the corresponding charged lepton can easily get a mass from
suppressed operators, while various other constraints are also
satisfied).  As a result, a large region of parameter space of the
model is still viable, and leads to a rather rich and interesting
phenomenology.

The most important features of such a framework include i) ``Dirac"
gauginos (especially gluinos) due to ``$R$-symmetry in the gauge
sector", ii) Absence of certain scalar trilinear terms and the
``standard" $\mu$-parameter (that we have called $\hat{\mu}$ in this
work) due to the ``$R$-symmetry in the matter sector", iii) Existence
of RPV operators of the type $\lambda_{ijk} L_i L_j E^c_k$,
$\lambda'_{ijk} L_i Q_j D^c_k$ and $B\mu_L^{(i)} H_u L_i$ consistent
with the leptonic $R$-symmetry $R=R_1$.  Due to a significant
sneutrino {\it vev}, the sneutrino acts as a Higgs field providing the
down-type fermion masses through subsets of these operators, iv)
Mixing between neutralinos and neutrinos, and between charginos and
charged leptons.  These features combine to give a rich and complex
pattern of signals at the LHC, which is studied in a companion paper
\cite{Frugiuele:2012xx}.  Although subsets of these features are
shared by other models, the entire pattern of signals is rather unique
and should be distinguishable from other models.  Here we have
highlighted how the (perhaps imminent) observation of lepto-quark
signals could be construed as a powerful indication that the TeV scale
Lagrangian indeed displays an approximate $U(1)_R$ symmetry of the
leptonic type.  We have also discussed how certain topologies at the LHC
could be used to infer that the sneutrino \textit{vev} is
non-vanishing (and large), and also that a missing energy observation
should be interpreted as being associated with a neutrino (as opposed
to a neutralino or a gravitino).  Further exploration of these issues
will certainly be welcome in the near future.

We emphasize that these models can easily accommodate the
observed $\sim 125$~GeV Higgs-like signal with significantly less tuning
than in the MSSM. The main ingredient is the existence of a scalar
triplet which must be somewhat heavy ($\gtrsim 1~{\rm TeV}$), as
required by current constraints on the $\rho$-parameter.  If such a
triplet scalar has an order one coupling to the Higgs, it can
contribute significantly to the Higgs mass at the loop level (the
$R$-symmetry forbids other tree-level contributions).  This is similar
in spirit to using the stops for such effect within the MSSM, but with
significantly less tuning than in that well-known case.  This observation is more general than the particular model we are
considering, but it fits rather nicely within the $U(1)_R$ framework.

Finally, from the theoretical point of view, it would be very
interesting to find a simple \emph{dynamical} mechanism of
supersymmetry breaking and mediation which preserves an approximate
$U(1)_R$ symmetry, followed by an $R$-symmetry breaking and mediation
mechanism, which generates the structure of operators as envisioned in
this class of models.

\vspace{0.4cm}

{\bf Acknowledgments}

C.F. and T.G. are supported in part by the Natural Sciences and
Engineering Research Council of Canada (NSERC).  E.P. is supported by
the DOE grant DE-FG02-92ER40699.  P.K. has been supported by the DOE
grant DE-FG02-92ER40699 and the DOE grant DE-FG02-92ER40704 during the
course of this work.

\appendix

\section{$R$-breaking Operators}
\label{r-breakops}

As explained in Section~\ref{r-break}, there are two natural ways in
which $R$-breaking can be transmitted to the visible sector: i)
generic gravity-mediation, and ii) anomaly-mediation.

Starting with generic gravity-mediation, we write \textit{arbitrary
Planck suppressed} couplings that are restricted only by the SM gauge
symmetries, but which do not respect the (anomalous) $U(1)_{R}$
symmetry (for concreteness, here we consider the case $R=R_1$).
Nevertheless, we assume that $M_{\star} \ll M_{P}$, so that the
$U(1)_{R}$ symmetry is approximate in the observable sector, being
dominated by the physics at $M_{\star}$.  Thus, we have
$U(1)_{R}$-violating superpotential terms:
\ba
\hspace{-3mm}
W_{\cancel{R}} &=& \hat{\mu} H_u H_d + \mu'_{i} H_{u} L_{i}   + \frac{1}{2} M_{S} S^2 + \frac{1}{2} M_{T} T^{2} + \frac{1}{2} M_{O} O^{2}
+ \tilde{y}^{d}_{ij} H_{d} Q_{i} D^{c}_{j} + \tilde{y}^{e}_{ij} H_{d} L_{i} E^{c}_{j} + \frac{1}{3} \, y_{S} S^{3},
\label{WRbreaking}
\ea
which can schematically arise from 
\ba
\label{yuk-rbreak}
&& \hspace{-1cm}
\int \! d^{4} \theta \, \frac{X^{\dagger}}{M_{P}} \left\{ H_u H_d + 
H_{u} L_{i} +  S^2 + T^{2} +O^{2} + \frac{1}{M_{P}} (H_{d} Q_{i} D^{c}_{j} + H_{d} L_{i} E^{c}_{j} + S^{3})
\right\}~.
\ea
We see that $\mu'_{i} \sim F_{X}/M_{P} \sim m_{3/2}$.  In this case,
we also see that the contribution to the ``standard'' (but
$U(1)_{R}$-violating) down-type and lepton Yukawa couplings is
$\tilde{y}^{d}_{ij} \sim \tilde{y}^{e}_{ij} \sim m_{3/2}/M_{P} <
10^{-22}$ for $m_{3/2} \lesssim {\rm MeV}$.  A similar suppression can
be expected for the $U^{c}D^{c}D^{c}$ superpotential operator and
other $R$-violating trilinears.  There are also $U(1)_{R}$-violating
soft-breaking terms:
\ba
V^{\rm soft}_{\cancel{R}} &=& B {\mu}_R H_{u} R_{d} +  A^{u}_{ij} H_{u} Q_{i} U^{c}_{j}
+ A^{d}_{ij} H_{d} Q_{i} D^{c}_{j} + A^{e}_{ij} H_{d} L_{i} E^{c}_{j}  + \tilde{A}^{d}_{ijk} L_{i} Q_{j} D^{c}_{k} + \tilde{A}^{e}_{ijk} L_{i} L_{j} E^{c}_{k}~,
\ea
which can arise from 
\ba
\label{soft-rbreak}
\int \! d^{4} \theta \, \frac{X^{\dagger}X}{M_{P}^{2}} \left\{ 
H_{u} R_{d} + \frac{1}{M_{P}} \left[ H_{u} Q_{i} U^{c}_{j} + \cdots \right]
\right\}~,
\ea
giving contributions to the $b$-terms of order $m_{3/2}^2$, and very
suppressed contributions to the $A$-terms, of order $(m_{3/2}/M_{P}) \,
m_{3/2}$.  However, there are also terms of the form
\ba
\label{trilinear}
\int \! d^{2} \theta \, \frac{X}{M_{P}} \left[ H_{u} Q_{i} U^{c}_{j} + \cdots \right]~,
\ea
that give contributions to the $A$-terms of order $m_{3/2}$.  Again,
we consider only Planck suppressed couplings between the hidden and
observable fields, so that no $\int \!  d^2\theta \, X H_{u} H_{d}$
operator, which would induce a too large $B {\mu}_R$ term, exists.
Similarly, there can exist $U(1)_{R}$-violating gaugino Majorana
masses $\frac{1}{2} M_{a} \lambda^{a}\lambda^{a}$, induced by
\ba
\label{Majorana}
\frac{1}{2} \int \! d^{2}\theta \frac{X}{M_{P}} {\cal W}^{\alpha} {\cal W}_{\alpha} + {\rm h.c.}
\ea
so that $M_{a} \sim m_{3/2}$.  Thus, the scale of
$U(1)_{R}$-preserving terms is taken to be $M_{\rm SUSY}$, which is
assumed to be much larger than the scale of $U(1)_{R}$-violating
operators, set by $m_{3/2}$.  All possible dimensionful terms allowed
by the gauge symmetries are allowed, and are induced at least at order
$m_{3/2}$.  Note that singlets in theories of the present type have
been argued to be safe from radiatively-induced
tadpoles~\cite{Goodsell:2012fm}.  Finally, as mentioned in the main
text, the experimental upper bound on neutrino masses puts an upper
bound on $m_{3/2}$ in this case to be ${\cal O}({\rm keV})$
\cite{Bertuzzo:2012su}.

In the case of anomaly-mediation, we imagine a situation in which the
``tree-level" transmission of $R$-breaking by operators
(\ref{yuk-rbreak}), (\ref{soft-rbreak}), (\ref{trilinear}) and
(\ref{Majorana}) is very suppressed, e.g.~due to sequestering.  Then
anomaly-mediation generates essentially the same operators
(\ref{trilinear}) and (\ref{Majorana}), except that they are generated
at one-loop rather than at tree-level.  So, the coefficients above are
suppressed by the one-loop factor $\sim \frac{1}{16\pi^2}$, giving
rise to an associated mass scale:
\ba 
M_a^{\rm anomaly}, A^{\rm anomaly} &\sim& \frac{m_{3/2}}{16 \pi^2}~. 
\ea 
In this case, the upper bound on $m_{3/2}$ from the neutrino mass
constraint is ${\cal O}(1-100)\,{\rm MeV}$ \cite{Bertuzzo:2012su}.

\section{A Flavor Ansatz}
\label{ansatz}

In Section~\ref{llp-const}, we saw that there exist a number of
constraints on the $\lambda$ and $\lambda'$ couplings from various
flavor-violating processes.  In Section~\ref{pheno}, we also studied
phenomenological consequences of an ansatz in which all the $\lambda$
and $\lambda'$ couplings are assumed to saturate the bounds in Tables
\ref{lambda-bounds}, \ref{lambdap2-bounds} and \ref{lambdap3-bounds}
(except $\lambda'_{311}$, which is taken to be negligible.).  However,
it is also possible to consider other simple ans\"atze about the
flavor dependence of these couplings, and we discuss such an option
here which allows a nice understanding of the relative magnitudes of
the various couplings and also satisfies existing constraints.  Also,
it has the advantage of \emph{further} reducing the number of
independent $\lambda$ and $\lambda'$ couplings.

We imagine that the model described by the superpotential in
(\ref{W-final}) has a flavor symmetry $G_F\equiv SU(3)_Q\times
SU(3)_U\times SU(3)_D\times SU(3)_L\times SU(3)_E$ at some high scale
which is broken by scalar components of ``flavon" superfields.  This
implies that the up-type Yukawa couplings and the $\lambda$ and
$\lambda'$ couplings (which include down-type and lepton Yukawa
couplings) should be thought of as flavon (super) fields transforming
in a non-trivial representation of $G_F$.  Of course, the symmetry
group $G_F$ is broken when their scalar components get ${\it vev}$'s
and give rise to the numerical values of the Yukawa couplings.  The
representations in Table \ref{yuk-repn} allow all the relevant
operators in the superpotential in (\ref{W-final}).
\begin{table}[t]
\vspace{-10pt}
\begin{center}
\begin{tabular}{|c|c|c|c|c|c|}
\hline
\rule{0mm}{5mm}
 & $SU(3)_Q$  & $SU(3)_U$ & $SU(3)_D$ & $SU(3)_L$ & $SU(3)_E$ \\
[0.3em]
\hline
$Q$ & $\tiny\yng(1)$ &  1 & 1 & 1 & 1   \\
[0.3em]
$U^c$ & 1 & $\tiny\yng(1)$  & 1 & 1 & 1   \\
[0.3em]
$D^c$ & 1 &  1 & $\tiny\yng(1)$  & 1 & 1   \\
[0.3em]
$L$ &  1 & 1 & 1 & $\tiny\yng(1)$ &  1    \\
[0.3em]
$E^c$ &  1 & 1 & 1 & 1 & $\tiny\yng(1)$   \\
[0.3em]
$Y_u$ & $\overline{\tiny\yng(1)}$ & $\overline{\tiny\yng(1)}$  & 1 & 1 & 1  \\
[0.3em]
$\lambda'$ & $\overline{\tiny\yng(1)}$ & 1 &  $\overline{\tiny\yng(1)}$ &   $\overline{\tiny\yng(1)}$  & 1    \\
[0.3em]
$\lambda$ &  1 & 1 & 1 & $\tiny\yng(2)$ & $\overline{\tiny\yng(1)}$   \\
[0.3em]
\hline
\end{tabular}
\end{center}
\vspace{-10pt}
\caption{\small{Representations of the fields and Yukawa spurions
under $G_F$.}}
\label{yuk-repn}
\vspace{-10pt}
\end{table}

In particular, the above choice gives rise to the following structure
for the coefficient of $L_i L_j E^c_k$:\footnote{In this section only,
we use subscripts (superscripts) to denote transformation in the
fundamental (anti-fundamental) of $SU(3)$.}
\ba 
\lambda^{ijk} = \epsilon^{ijn}\,Y_{n}^k~,
\ea 
where $\epsilon^{ijk}$ stands for the antisymmetric tensor with three
indices, and $Y_n^k$ transforms in the
$(\tiny\yng(1),\overline{\tiny\yng(1)})$ of $SU(3)_L\times SU(3)_E$.
Then, using the $SU(3)_L$ and $SU(3)_E$ rotation freedoms for $L_j$
and $E^c_k$, respectively, it is possible to choose a convenient form
of $Y_{n}^k$.  In particular, by choosing:\footnote{This can be
achieved by diagonalizing $Y_{n}^k$ by bi-unitary transformations,
followed by the exchange of the second and third rows, with an
appropriate sign.}
\ba 
\label{choice}Y_{n}^{k} &=& y_1\,\delta_{n}^{k}~,
\hspace{2cm} n=1~,
\nonumber\\
&=& -y_3\,\delta_{3}^{k}~,
\hspace{1.7cm}n=2~,
\nonumber\\
&=& y_2\,\delta_{2}^{k}~,
\hspace{2.04cm}n=3~,
\ea 
one gets a simple structure of the $\lambda_{ijk}$ couplings, in which
the matrix $\lambda_{1jk}$ is a diagonal matrix corresponding to the
usual leptonic Yukawa couplings appearing in $W_{{\rm Yukawa}}$ in
Eq.~(\ref{convbasis}) with $y_2=y_2^{(e)}\equiv y_{\mu}$, and
$y_3=y_3^{(e)}\equiv y_{\tau}$.\footnote{Note the electron mass arises
not from superpotential couplings, but from SUSY breaking operators;
see Section~\ref{fermionic-ew}.}  Furthermore, the matrices
$\lambda_{2jk}$ and $\lambda_{3jk}$ have only one independent
$\lambda_{ijk}$ coupling: $\lambda_{231}=-\lambda_{321} = y_1$; the
rest either vanish or are related to the previous Yukawa couplings.

The $L Q D^c$ operators are less constrained by these arguments.
However, it is convenient to make the following simple ansatz:
\ba
\label{ansatz-lambdap} 
(\lambda')^{ijk} \sim y_i\, (\hat{Y}_d)^{jk}~,
\ea 
where $y_i$ are defined in (\ref{choice}), and $(\hat{Y}_d)^{jk}$
transforming in the
$(\overline{\tiny\yng(1)},\overline{\tiny\yng(1)})$ of $SU(3)_Q\times
SU(3)_D$.  Again, by choosing an appropriate $SU(3)_Q \times SU(3)_D$
rotation, $(\hat{Y}_d)^{jk}$ can be brought into the form ${\rm diag}
(\hat{y}_1,\hat{y}_2,\hat{y}_3)$, which results in a very simple
structure of the $\lambda'_{ijk}$ couplings as well.  In particular,
the matrix $\lambda'_{1jk}$ is diagonal and corresponds to the usual
down-type quark Yukawa couplings in $W_{{\rm Yukawa}}$ in
(\ref{convbasis}).  The remaining $\lambda'$ couplings are
\emph{completely} fixed by these down-type Yukawa couplings and the
numbers $\{y_1,y_2,y_3\}$ in (\ref{choice}).  Since $y_2$ and $y_3$
are the muon and tau Yukawa couplings, the only other independent
input to fix all the remaining ones is $y_1 =
\lambda_{231}=-\lambda_{321}$.

Since the above ansatz provides a concrete determination of all the
couplings, it is straightforward (but tedious) to check constraints
from all combinations of $\lambda$ and $\lambda'$ couplings in
\cite{Barbier:2004ez, Saha:2002kt, Dreiner:2006gu, Dreiner:2012mx}.
We find that all such constraints are readily satisfied, except that
coming from $|\lambda_{231}\,\lambda'_{311}|$, which is violated by a
small amount.  This, however, just means that
(\ref{ansatz-lambdap}) is only approximate and receives some
corrections.  In any case, the main motivation for the ansatz
(\ref{ansatz-lambdap}) is to simplify the structure of the $\lambda'$
couplings, and perhaps suggest a possible high-energy rationale for their
structure.

Finally, since the only new coupling within this approach is $y_1$, it
is important to know what one would generically expect for its size.
In fact, the bound on the product $|\lambda'_{133}\,\lambda'_{233}|
\equiv y_b\,|\lambda'_{233}| < 1.1\times 10^{-5}$~\cite{Dreiner:2012mx}
puts a bound on $y_1$, since in our framework $y_b\,\lambda'_{233}
\simeq \frac{y_{\mu}\,y_b^2}{y_1} \simeq (1.53\times
10^{-7}\,\sec^3\beta) / y_1$.  Therefore, the bound implies that the
\emph{largest} non-Yukawa $\lambda'$ coupling ($\lambda'_{333}$) is
roughly given by:
\ba 
\lambda'_{333} &\sim& \frac{y_{\tau}y_b}{y_1} ~\lesssim~ 0.011\,\cos\,\beta~,
\ea 
which is quite small.  However, this bound is specific to the previous
ansatz and, experimentally, $\lambda'_{333}$ can be significantly
larger, as described in the main text (where we do not use the ansatz
of this section).  We still describe it here as it is simple and will
also be considered in our LHC phenomenology study
in~\cite{Frugiuele:2012xx}.

\section{Lower Bound on $\lambda'_{i33}$ given the Observation of an LQ Signal}
\label{lambdap-bound}

In this appendix we argue that typically the size of the relevant
$\lambda'$ couplings cannot be extremely suppressed, if a LQ signal
was observed at the LHC. In order to estimate a \textit{lower} bound
on $(\lambda')^2$ such that the BR in the LQ channel is not very
suppressed (and could be observed in the near future), we explore
several possibilities, depending on whether the lepto-quark is
$\tilde{t}_L$, $\tilde{b}_L$ or $\tilde{b}_R$.  To be definite we
estimate the ``minimum" value of $(\lambda')^2$ by requiring that the
partial decay width in the lepto-quark channel, given by $\Gamma_{\rm
LQ} = [(\lambda')^2/16\pi] \, m_{\rm LQ}$ equals one of the standard
R-parity conserving 2-body decay widths~\cite{Djouadi:2000bx},
$\Gamma_{\rm 2-body}$, interpreted within the MSSM:~\footnote{We show
the results within the MSSM structure to illustrate what an
interpretation outside the $U(1)_R$ symmetric framework would entail.
A similar analysis within the $L=R$ model results in qualitatively
similar features regarding the expected sizes of the $\lambda'$
couplings.}
\begin{figure}[t]
\vspace{-5mm}
\centering
\includegraphics[width=0.76\textwidth]{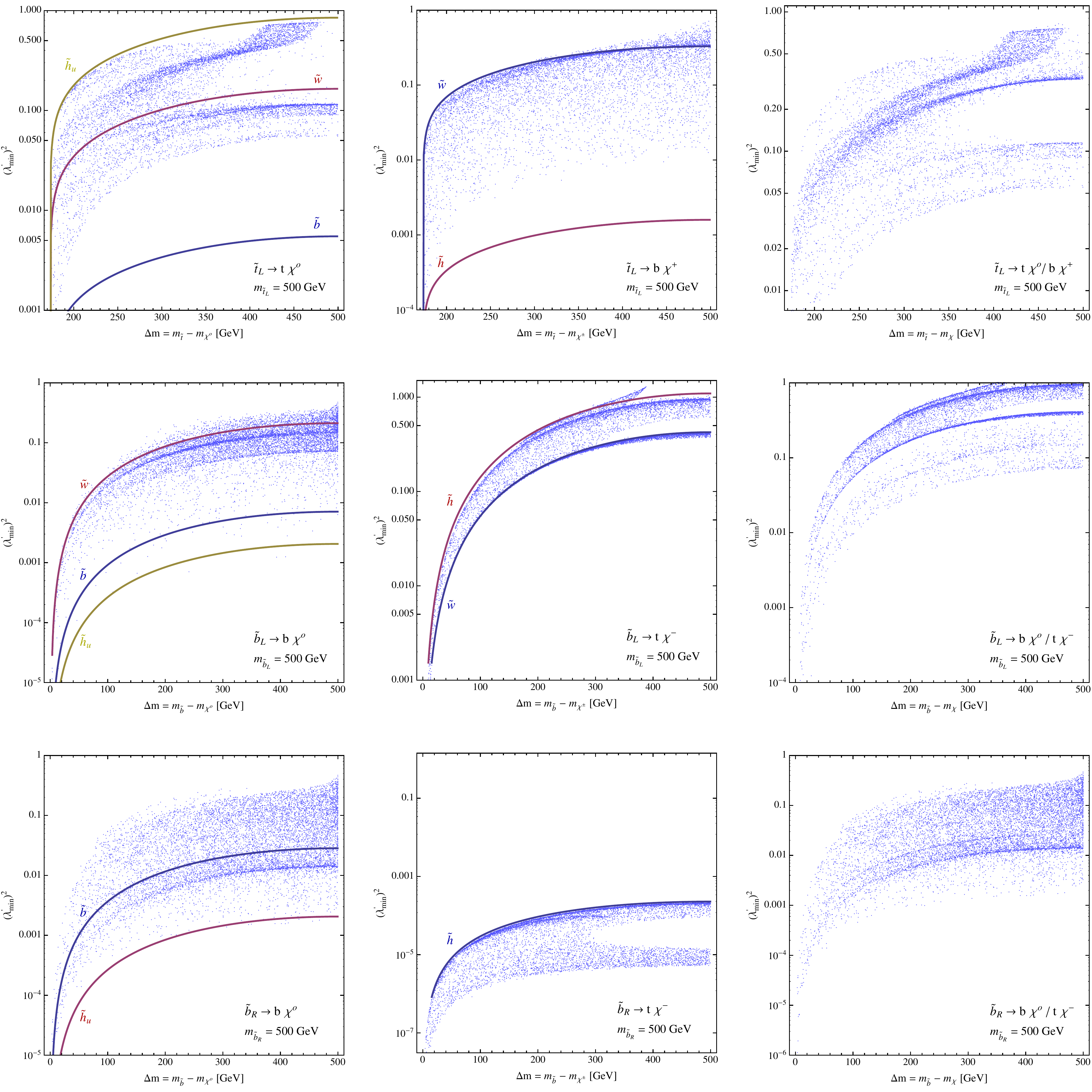}
\caption{\footnotesize{Required value of $(\lambda')^2$ such that the
partial width in the lepto-quark signal equals the standard 2-body
decay width into neutralinos/charginos for a $500~{\rm GeV}$
$\tilde{t}_L$ (upper row), $\tilde{b}_L$ (middle row) and
$\tilde{b}_R$ (lower row), within the RPV-MSSM. The third column uses
the largest of the neutralino/chargino channels to bound
$(\lambda')^2$ for any given point in the random scan over $M_1$,
$M_2$, $\mu$ and $\tan\beta$.  The smooth curves indicate the limit
where the neutralino is pure bino, pure wino and pure higgsino, as
indicated, and similarly for the chargino being pure wino or higgsino.}}
\label{fig:lepto-quarkCouplingBoundMSSM}
\end{figure}
\ba
(\lambda'_{\rm min})^2 &\equiv& \frac{16\pi}{m_{\rm LQ}} \, \Gamma_{\textrm{2-body}}~.
\ea
We show in Fig.~\ref{fig:lepto-quarkCouplingBoundMSSM} the result for
$500~{\rm GeV}$ lepto-quarks, when comparing to their partial decay
widths into a neutralino plus quark (left column) and into a chargino
plus quark (middle column).  We have performed a scan over $M_1$,
$M_2$, $|\mu|$ $\in [0,600]~{\rm GeV}$ and $\tan\beta \in [3,50]$,
diagonalizing the MSSM neutralino and chargino mass matrices to find
the spectrum and composition of the eigenstates for each parameter
point.  We compare to the \textit{dominant} neutralino or chargino
channel, and plot the $(\lambda'_{\rm min})^2$ defined above, as a
function of $\Delta m = m_{\rm LQ} - m_{\chi}$, where $m_{\rm LQ} =
500~{\rm GeV}$ is the lepto-quark mass and $m_{\chi}$ is the
appropriate neutralino or chargino mass.  We also show curves
corresponding to the limiting cases in which the neutralino is pure
bino, pure wino or pure Higgsino ($\tilde{h}_u$), and also when the
chargino is pure gaugino or pure Higgsino.  For each scanned parameter
point we have also estimated $(\lambda'_{\rm min})^2$ based on the
\textit{largest} partial decay width of \textit{any} of the neutralino
\textit{and} chargino channels (shown in the right column plots).  We
will use the latter as our estimate for $(\lambda'_{\rm min})^2$.

We see from the plots in the right column that in the case that the
lepto-quark is a $SU(2)_L$ doublet, $(\lambda'_{\rm min})^2$ is
typically above $0.01$ (unless the decay is very close to threshold).
We also note that the bulk of the cases has an even larger
$(\lambda'_{\rm min})^2 \gtrsim 0.1$.  Our estimate given in
(\ref{LR-upper}) is obtained by using $(\lambda'_{\rm min})^2 \sim
0.01-0.1$ in (\ref{AbBound}), together with $\Delta m_\nu \sim
0.1~{\rm eV}$.  We also see from the lower row in
Fig.~\ref{fig:lepto-quarkCouplingBoundMSSM} that when the lepto-quark
is $\tilde{b}_R$, our estimate for $(\lambda'_{\rm min})^2$ is weaker,
although still larger than what the neutrino bound requires if there
was no suppression from LR mixing in the sbottom sector.  These two
cases could be distinguished by the type of lepto-quark signal: $b l$
for the doublet (we expect the mass splitting between $\tilde{t}_L$
and $\tilde{b}_L$ to not exceed a few tens of GeV), and $t l$ for
the singlet.  Both cases should be accompanied by a $b$ +
$\slash{\!\!\!\!E}_T$ signal, that would suggest that the missing
energy comes from a neutrino.

\bibliographystyle{utphys}
\bibliography{Paper1}

\end{document}